\documentclass[version1996]{siggraph}
\usepackage{epsf}

\def\s{\sigma}

\def\half{{\textstyle{{1}\over{2}}}}
\def\df{\partial}
\def\nn{\nonumber}

\def\Z{{\bf Z}}
\def\R{{\bf R}}

\def\mod{\mathop{\mbox{mod}}\nolimits}

\def\ket#1{\bigl|#1\bigr>}
\def\bra#1{\bigl<#1\bigr|}
\def\vsp{\vspace{0mm}}
\def\ll{\mathop{<\!<}}
\def\gg{\mathop{>\!>}}
\newtheorem{lemma}{Lemma}

\newlength{\mywidth}\mywidth=2.3truein 
\epsfysize=\mywidth
\epsfxsize=\mywidth
\def\fref#1{fig.\ref{#1}}

\renewenvironment{figure}{\refstepcounter{figure}
\baselineskip=0.4\normalbaselineskip\footnotesize}
{\baselineskip=\normalbaselineskip}
\def\fignum{{\bf Fig.\arabic{figure}.\quad}}
\renewenvironment{table}{\refstepcounter{table}}{}
\def\tabnum{Table \arabic{table}}

\begin{document}

\title{String theory in Lorentz-invariant light cone gauge - III}
\author{Igor Nikitin, Lialia Nikitina\\
\\
{\it Fraunhofer Society, IMK.VE, 53754, St.Augustin, Germany}
}
\date{}
\maketitle

\quad

\vspace{-10mm}

This paper completes the work, initiated in \cite{partI,partII}, 
further referred as Parts I and II, concerning to Dirac's quantization 
of Nambu-Goto theory of open string, formulated in the space-time 
of dimension $d=4$. Here we perform more detailed study of Gribov's copies 
in the classical mechanics and determine the quantum spectrum 
of masses for the arbitrary spin case.


\section{Gribov's copies}

Gribov's copies are multiple intersections of the orbit 
of gauge group and the surface of gauge fixing condition.
In Part~II we have shown that the considered version of string
theory possesses Gribov's copies, related with the singular points 
of a specially constructed vector field on the sphere. 
Therefore, the following topological invariants can be introduced 
for Gribov's copies.  

\vspace{2mm}
\noindent{\it Definition 1:} let the phase space $M$ be 
a smooth orientable manifold. Let the orbit of gauge group $G$ 
and the surface of gauge fixing condition $F$ be its smooth 
orientable submanifolds with $dim(G)=codim(F)$. 
Let $P$ be the point of their transversal intersection.
This means that in point $P$ the tangent spaces to $G$ and $F$ 
span the tangent space to $M$. Let $\vec\tau(F,P)$,
$\vec\tau(G,P)$ and $\vec\tau(M,P)$ be the bases 
in the tangent spaces, defining the orientation of $F$, $G$ and $M$,
evaluated in point $P$. {\it The index of intersection} 
of $F$ and $G$ in point $P$ is defined as a number $\nu$,
equal to $(+1)$ if the basis $(\vec\tau(F,P),\vec\tau(G,P))$ 
has the same orientation as the basis $\vec\tau(M,P)$,
and equal to $(-1)$ if the orientations are opposite.

\vspace{2mm}
\noindent{\it Definition 2:} non-degenerate singular points 
of 2-dimensional vector field are focus, center, node and saddle.
Standard topological classification introduces {\it the index 
of singular point} $\nu$, equal to $(+1)$ for the cases 
(focus, center, node) and equal to $(-1)$ for the saddle.

\begin{lemma}\label{Lgrib1} (topological charges of Gribov's copies):\rm\\
${}$ ~ \hspace{12mm} ~ these two definitions coincide.
\end{lemma}

The explicit expressions for the vector fields 
are given by Eq.(12) in Part~II. There is also an alternative definition:
\begin{eqnarray}
&&~\hspace{-10mm}
{\vec a}_{n}={\textstyle{{1}\over{2\sqrt{2\pi}}}}
\oint d\vec Q(\s)\cdot\nn\\
&&\cdot\exp\left[
{\textstyle{{2\pi in}\over{L_{t}}}}
\left(L(\s)-(\vec Q(\s)-\vec X)\vec e_{3}\right)\right],
\label{oint2}
\end{eqnarray}
where $\vec X=\oint dL(\s)\vec Q(\s)/L_{t}$ 
defines an average position of the curve $\vec Q(\s)$
and coincides with the definition of mean coordinate
in CMF given in Part~I. The difference of (\ref{oint2})
and PII-Eq.(12) consists in $\vec e_{3}$-dependent 
phase factor $e^{in\varphi(\vec e_{3})}$, $\varphi(\vec e_{3})=
{2\pi}(\vec X-\vec Q(0))\vec e_{3}/{L_{t}}$.
According to the Part~II, the evolution of ${\vec a}_{n}$ 
is the phase rotation. Thus, the phase factor actually introduces 
a difference of ``local time'' for the evolution of vector fields
on the sphere.

\newpage ~

\vspace{-13mm}

\begin{lemma}\label{Lgrib_ph} (phase factor):\rm 
~~the factor $e^{in\varphi(\vec e_{3})}$ preserves the orbits of 
Gribov's copies and changes their evolution parameter
from lcg's PII-Eq.(13) to the natural one (length of the curve).
\end{lemma}

Further we will use the definition (\ref{oint2}),
more convenient to study the evolution of Gribov's copies.

\vspace{-4mm}


\begin{figure}\label{fgrib}
\begin{center}
~\epsfxsize=9cm\epsfysize=13.5cm\epsffile{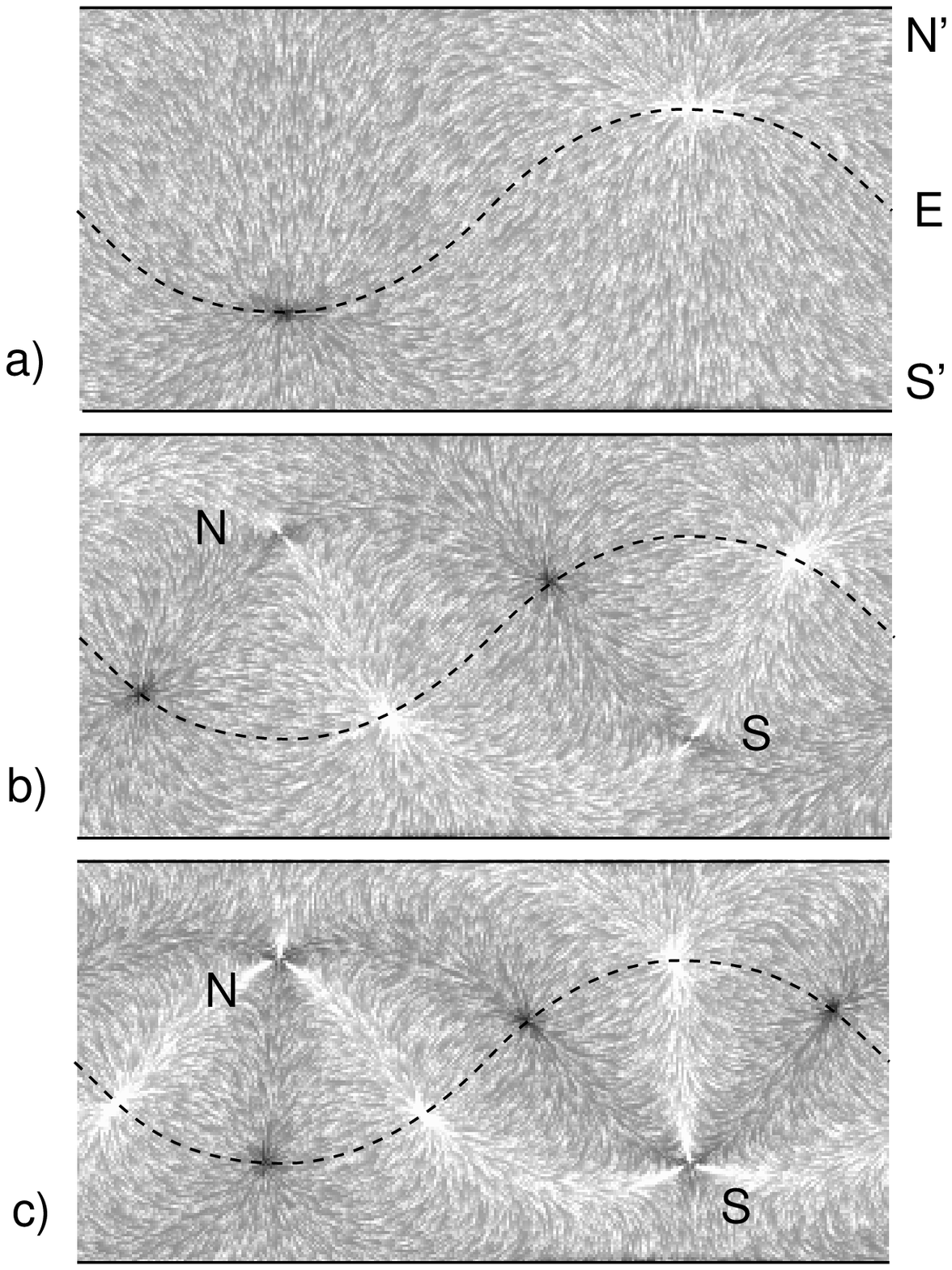}
\end{center}

\vspace{-2mm}

\fignum Gribov's copies. These images are created by a computer program for
the visualization of vector fields, courtesy of Wasil Urazmetov, 
IHEP, Protvino, Russia. A paper on the visualization techniques used
in this program will be published elsewhere.
\end{figure}

\vspace{2mm}

To investigate the properties of Gribov's copies in more detail, 
we perform their visualization. The vector fields of \fref{fgrib}
are displayed on a spherical map (longitude, latitude), where 
the top and bottom lines correspond to auxiliary poles N',S'.
These poles are specially selected in regular points of the vector field,
to prevent the image distortion in the vicinity of the original poles 
N,S, where the singularities are located. The equator E for the poles N,S
is shown on the images by dashed line. 
The following singularities of the vector field 
are visible in these images: nodes -- as spots of black/white color, 
dependent on the field direction (source/sink); 
saddles -- as cross-shaped black-white spots with a hyperbolic 
structure of the field in the vicinity. The visualization allows to
formulate the following lemma (analytically proven in Appendix 1).

\begin{lemma}\label{Lgrib2} 
(nearly straight solution, analog of PII-L5,6):\rm\\
For the straight line string the singular points 
of the vector field $\vec q_{s\perp}(\vec e_{3})$ are:
at $s=1$ two nodes on the equator (\fref{fgrib}a),
at $s=2$ saddles on the northern and southern poles
and four nodes on the equator (\fref{fgrib}b),
at $s>2$ multi-saddles on the northern and southern poles
and $2s$ nodes on the equator ($s=3$, \fref{fgrib}c).
During the evolution the nodes move along the equator,
while the singular points in the poles stay fixed
and saddle patterns rotate around them. After a small deformation 
of the string from straight configuration the nodes
move along a common trajectory in the vicinity of the equator;
at $s=2$ the saddles move in small loops near the poles; at $s>2$ 
the multi-saddles are unfolded to $(s-1)$ non-degenerate saddles
moving near the poles. After the lapse of time 
$\Delta\tau_{s}=\pi/s=$(period of evolution)$/2s$ 
the vector field reverses its direction and the pattern of 
singularities returns to the initial state. 
During this time the equatorial singularities move to the neighbor ones, 
pole singularities at $s=2$ perform one revolution along the loops.
\end{lemma}

The discussion of the properties of Gribov's copies
in quantum theory can be found in Appendix~3.

\section{Arbitrary spin case}
\subsection{Classical mechanics}
Let's define $\gamma=(L_{0}^{(2)})^{-1/2}$.

\begin{lemma}\label{Lstruct} (structural, analog of PII-L2): \rm
$a_{2}$ expansion in the vicinity of straight-line string has a form
\begin{eqnarray}
&&a_{2}=\sum\limits_{n\geq1}{P_{n}a_{1}^{2}\over|a_{1}|^{4n}},\label{a2cl}
\end{eqnarray}
where $P_{n}$ are polynomials of $a_{1},\Sigma_{-},S_{-},d,f,g'$,
their conjugates and $\gamma,\gamma^{-1}$. The polynomials possess 
the following properties: (i) each monomial in $P_{n}$ contains, 
counting powers, $n$ variables 
from the group ($\Sigma_{-},S_{-},d,f,g'$ and their conjugates) and
$2(n-1)$ variables from the group ($a_{1},a_{1}^{*}$);
(ii) each monomial contains at least one variable from the group
$(\Sigma_{\pm},S_{\pm})$; (iii) variables $a_{1}$ and $a_{1}^{*}$
enter in even powers only; (iv) each monomial, containing $S_{-}$,
also contains $(a_{1}^{*})^{2k}$ with $k\geq1$; 
(v) $P_{n}=O(\gamma^{-[(n+1)/2]})$ at $\gamma\to0$.
\end{lemma}

For the purposes of further consideration it is convenient 
to extract from $a_{2}$ a common phase multiplier 
$a_{1}^{2}/|a_{1}|^{2}$ and to define
$a_{2}=\alpha_{2}\cdot a_{1}^{2}/|a_{1}|^{2}$, so that
$|a_{2}|=|\alpha_{2}|$ and

\begin{eqnarray}
&&\alpha_{2}=\sum\limits_{n\geq1}{P_{n}\over|a_{1}|^{4n-2}},\quad
{P^{2}\over2\pi}=L_{0}^{(2)}+2|\alpha_{2}|^{2}.\label{alp2cl}
\end{eqnarray}

Explicit expressions for the first three polynomials $P_{n}$ are:

\baselineskip=0.4\normalbaselineskip\footnotesize

\def\myfrac#1#2{#1/#2}

\vspace{2mm}\noindent
$P_{1}=-{\Sigma_{+}} + \myfrac{{S_{+}}}{{\gamma}},$

\vspace{2mm}\noindent
$P_{2}={{a_{1}^{*}}}^2\,{f^{*}}\,{\Sigma_{-}} + 
  {{a_{1}}}^2\,{{g'}^{*}}\,{\Sigma_{+}} - 
  \myfrac{{{a_{1}^{*}}}^2\,{f^{*}}\,{S_{-}}}
   {{\gamma}} - \myfrac{{{a_{1}}}^2\,{{g'}^{*}}\,
     {S_{+}}}{{\gamma}},$

\vspace{2mm}\noindent
$P_{3}=- {{a_{1}^{*}}}^4\,{f^{*}}\,{g'}\,
     {\Sigma_{-}}   - 
  {{a_{1}}}^2\,{{a_{1}^{*}}}^2\,{f^{*}}\,{{g'}^{*}}\,
   {\Sigma_{-}} - {{a_{1}^{*}}}^4\,{d^{*}}\,
   {{\Sigma_{-}}}^2 $

$- {{a_{1}}}^2\,{{a_{1}^{*}}}^2\,f\,
   {f^{*}}\,{\Sigma_{+}} - 
  {{a_{1}}}^4\,{{{g'}^{*}}}^2\,{\Sigma_{+}} - 
  \myfrac{{{a_{1}}}^2\,{{a_{1}^{*}}}^2\,d\,{\Sigma_{-}}\,
     {\Sigma_{+}}}{2} $

$+ \myfrac{{{a_{1}^{*}}}^4\,{f^{*}}\,
     {g'}\,{S_{-}}}{{\gamma}} + 
  \myfrac{{{a_{1}}}^2\,{{a_{1}^{*}}}^2\,{f^{*}}\,
     {{g'}^{*}}\,{S_{-}}}{{\gamma}} + 
  \myfrac{2\,{{a_{1}^{*}}}^4\,{d^{*}}\,{\Sigma_{-}}\,
     {S_{-}}}{{\gamma}} $

$+ \myfrac{{{a_{1}}}^2\,{{a_{1}^{*}}}^2\,d\,{\Sigma_{+}}\,
     {S_{-}}}{2\,{\gamma}} - 
  \myfrac{{{a_{1}^{*}}}^4\,{d^{*}}\,{{S_{-}}}^2}
   {{{\gamma}}^2} + \myfrac{{{a_{1}}}^2\,{{a_{1}^{*}}}^2\,
     f\,{f^{*}}\,{S_{+}}}{{\gamma}} $

$+ \myfrac{{{a_{1}}}^4\,{{{g'}^{*}}}^2\,{S_{+}}}
   {{\gamma}} + \myfrac{{{a_{1}}}^2\,{{a_{1}^{*}}}^2\,d\,
     {\Sigma_{-}}\,{S_{+}}}{2\,{\gamma}} + 
  {{a_{1}}}^2\,{{a_{1}^{*}}}^2\,{\gamma}\,
   {\Sigma_{-}}\,{\Sigma_{+}}\,{S_{+}} $

$- \myfrac{{{a_{1}}}^2\,{{a_{1}^{*}}}^2\,d\,{S_{-}}\,
     {S_{+}}}{2\,{{\gamma}}^2} - 
  {{a_{1}}}^2\,{{a_{1}^{*}}}^2\,{\Sigma_{+}}\,
   {S_{-}}\,{S_{+}} - 
  {{a_{1}}}^2\,{{a_{1}^{*}}}^2\,{\Sigma_{-}}\,
   {{S_{+}}}^2 $

$+ \myfrac{{{a_{1}}}^2\,{{a_{1}^{*}}}^2\,
     {S_{-}}\,{{S_{+}}}^2}{{\gamma}}.$

\baselineskip=\normalbaselineskip\normalsize

\subsection{Quantum mechanics}

\paragraph*{Spin operator} was defined in Part II either by components
in external reference frame (upper index, $S^{i}$) or by components in
the special coordinate system $\vec e_{i}$, related with string dynamics
(lower index, $S_{i}$). The commutation relations:
\begin{eqnarray}
&&[S^{i},S^{j}]=i\epsilon_{ijk}S^{k},\ 
[S_{i},S_{j}]=-i\epsilon_{ijk}S_{k},\
[S^{i},S_{j}]=0\nn
\end{eqnarray}
correspond to raising/lowering operators
\begin{eqnarray}
&&S^{\pm}=S^{1}\pm iS^{2},\ 
[S^{3},S^{\pm}]=\pm S^{\pm},\ [S^{+},S^{-}]=2S^{3},\nn\\
&&S_{\pm}=S_{1}\pm iS_{2},\ 
[S_{3},S_{\pm}]=\mp S_{\pm},\ [S_{+},S_{-}]=-2S_{3},\nn
\end{eqnarray}
i.e. $S^{+}$ raises $S^{3}$ by $1$, $S^{-}$ lowers $S^{3}$ by $1$; 
$S_{+}$ lowers $S_{3}$ by $1$, $S_{-}$ raises $S_{3}$ by $1$.
In further mechanics only low-index operators $S_{i}$ will participate. 
In Part II the oscillator variables were composed to the
operators $\Sigma_{\pm},A_{3}^{(2)}$ with commutation relations
$[A_{3}^{(2)},\Sigma_{\pm}]=\pm \Sigma_{\pm}$, i.e. $\Sigma_{+}$ raises 
$A_{3}^{(2)}$ by $1$, $\Sigma_{-}$ lowers $A_{3}^{(2)}$ by $1$.
These operators define a constraint $\chi_{3}=S_{3}-A_{3}^{(2)}$,
representing a symmetry of the mechanics,
under which the state vectors are invariant: 
$\chi_{3}\ket{\Psi}=0$. We see that the
operators $\Sigma_{\pm}$ and $S_{\pm}$ have similar 
commutation relations with $\chi_{3}$: 
$[\chi_{3},\Sigma_{\pm}]=\mp \Sigma_{\pm}$,
$[\chi_{3},S_{\pm}]=\mp S_{\pm}$,
so that a linear combination of $\Sigma_{+}$ and $S_{+}$ 
lowers $\chi_{3}$ by $1$, while a linear combination of 
$\Sigma_{-}$ and $S_{-}$ raises $\chi_{3}$ by $1$. We will also say
that $\Sigma_{+}$ and $S_{+}$ have $\chi_{3}$-charge $(-1)$, while
$\Sigma_{-}$ and $S_{-}$ have $\chi_{3}$-charge $1$.
Similarly the concept of charges, introduced in Part II, can be 
extended to other operators from oscillator and spin parts of the mechanics.

Matrix elements of spin components 
do not depend on the representation of the algebra
and have well known form \cite{LLifsh}: 
\begin{eqnarray}
&&\bra{S(S_{3}-1)S^{3}}S_{+}\ket{SS_{3}S^{3}}=\sqrt{S(S+1)-S_{3}(S_{3}-1)},
\nn\\
&&\bra{S(S_{3}+1)S^{3}}S_{-}\ket{SS_{3}S^{3}}=\sqrt{S(S+1)-S_{3}(S_{3}+1)},
\nn
\end{eqnarray}
all other elements vanish. Concrete representation of quantum top
is described by Wigner's functions ${\cal D}^{S}_{S_{3}S^{3}}$ 
\cite{Wigner,LLifsh}.
Here $S$ characterizes the eigenvalue of Casimir
operator $S^{i}S^{i}=S_{i}S_{i}=S(S+1)$, commuting with
all spin components. $S$ is 
integer for single-valued representation of $SO(3)$ 
and half-integer for double-valued one. In Appendix~2
we show that in our problem only integer values of $S$ 
should be taken.


\paragraph*{Ordering rules.} 
We use the same ordering of elementary operators 
as defined in Part II, Table~1. In the products of spin components
$S_{\pm}$, commuting with elementary operators, we select the
ordering $S_{+}S_{-}$, so that $S_{3}$-raising operator
$S_{-}$ stands on the right and annulates the states
with maximal spin projection $S_{3}=S$. We remind that
classically $S_{3}=S=P^{2}/2\pi$ corresponds to one-modal
solution, straight line string, associated with the leading 
Regge trajectory. Quantum analog of (\ref{alp2cl}) 
is constructed as follows:
\begin{eqnarray}
&&\alpha_{2}=\sum\limits_{n\geq1}\tilde n_{1}^{-2n+1}:P_{n}:,\label{alp2q}
\end{eqnarray}
where $\tilde n_{1}=a_{1}^{+}a_{1}+c_{1}$. Here we introduce a constant term 
$c_{1}$, whose contribution vanishes on classical level
(in the limit of large occupation numbers $a_{1}^{+}a_{1}$),  
and add analogous terms in quantum definition 
$\gamma=(L_{0}^{(2)}+c_{2})^{-1/2}$ and in the definition
of mass-squared operator, which we fix as follows:
\begin{eqnarray}
&&{P^{2}\over2\pi}=L_{0}^{(2)}+2\alpha_{2}\alpha_{2}^{+}+c_{3}.\label{P2q}
\end{eqnarray}
Note that this definition differs from one used in Part II
for $S=0$ case. Later we will show that this definition
is preferable for the description of arbitrary spin case.

\begin{lemma}\label{Lconv} (convergence, analog of PII-L7):\rm\\
$\bra{L_{0}^{(2)}=N_{1},S} :P_{n}: \ket{L_{0}^{(2)}=N_{2},S}=0$, if \\
$n>\min\{1+(N_{1}+N_{2})/2,\ (4+2(N_{1}+N_{2})+4S)/5\}$.
\end{lemma}

\begin{figure}\label{elka-corrected}
\begin{center}
~\epsfxsize=6cm\epsfysize=9cm\epsffile{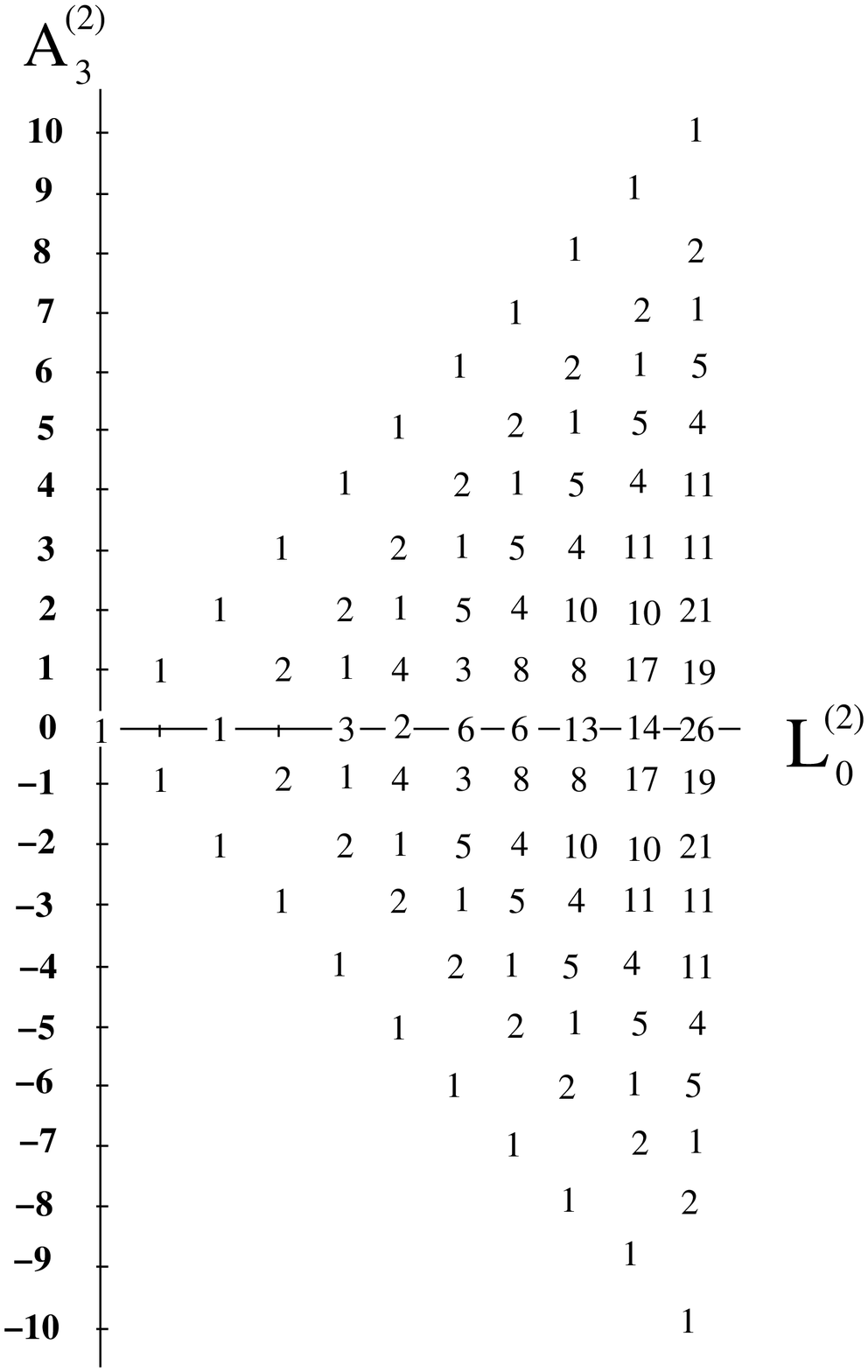}

\fignum Spectrum $(L_{0}^{(2)},A_{3}^{(2)})$.
\end{center}
\end{figure}

\begin{figure}\label{spec}
\begin{center}
~\epsfxsize=8cm\epsfysize=6cm\epsffile{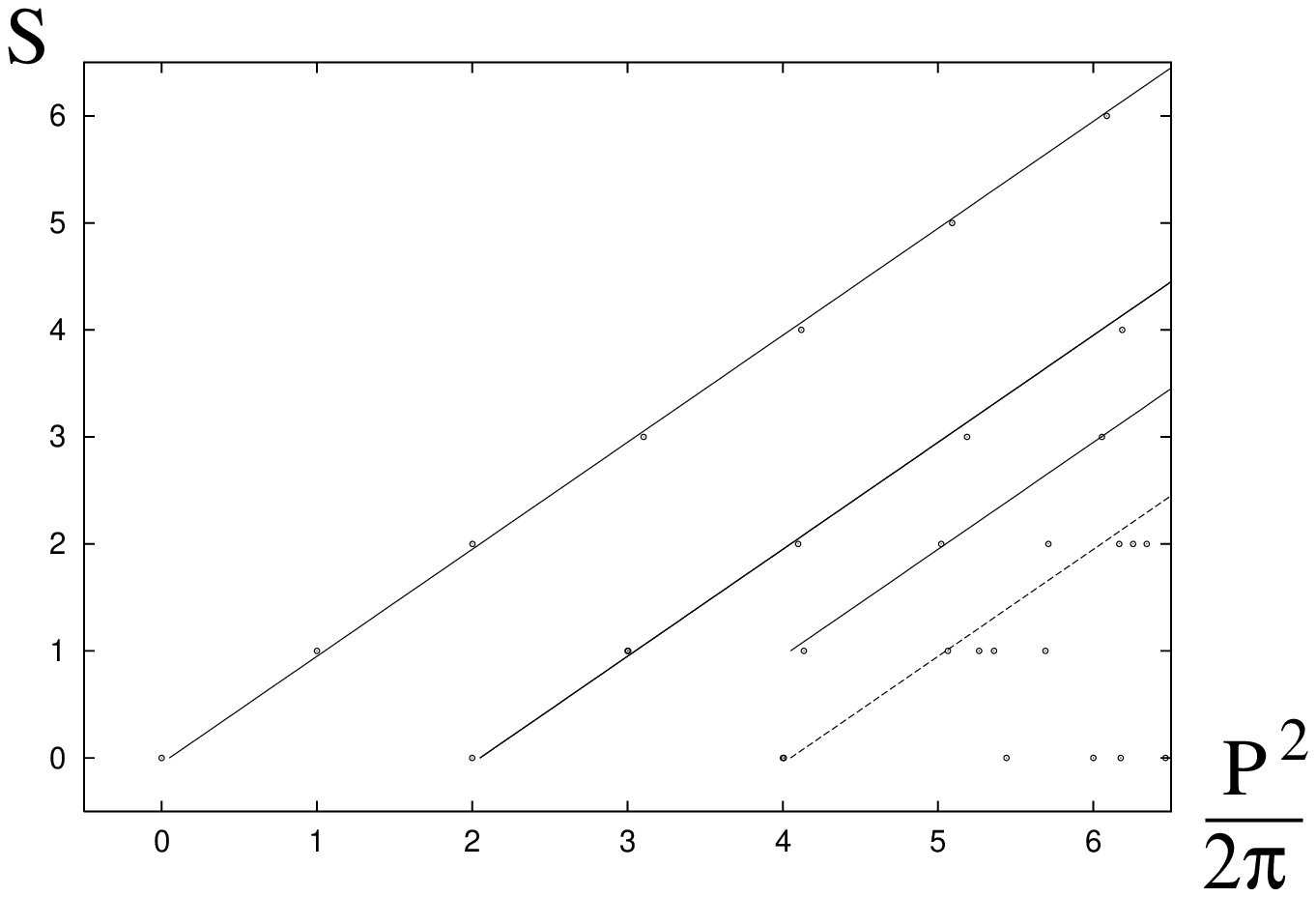}

\fignum Spectrum $(P^{2}/2\pi,S)$.
\end{center}
\end{figure}

The resulting spectrum $(P^{2}/2\pi,S)$ is shown on \fref{spec}.
The spectrum has common features with the upper part
of $(L_{0}^{(2)},A_{3}^{(2)})$ spectrum, 
shown on \fref{elka-corrected}. The beginning of the spectrum $(P^{2}/2\pi,S)$ 
consists of three almost linear Regge trajectories. There is a 2-unit 
gap between the first and the second trajectories. The third trajectory
starts at $S=1$ level. For the next trajectories the degenerate states of
\fref{elka-corrected} become splitted on \fref{spec}.
The states at $(P^{2}/2\pi,S)=(3,1)$ and $(4,0)$
comprise two numerically close pairs with $P^{2}/2\pi=3,3.0046$
and $P^{2}/2\pi=4,4.0066$, while the other states on \fref{spec}
are non-degenerate (not counting trivial degeneration for 
the upper-index $S^{3}=-S...S$ and direction of $P_{\mu}$). 
The spectrum is computed for the values
$c_{1}=2,\ c_{2}=4,\ c_{3}=0$. Smaller values of $c_{1},c_{2}$
correspond to higher non-linearities in the spectrum, while
larger values of $c_{1},c_{2}$ make the spectrum more linear
and closer to the spectrum of $(L_{0}^{(2)},A_{3}^{(2)}\geq0)$.
This behavior is characterized by the following lemmas,
where for clarity of statements we fix $c_{3}=0$.

\vspace{3mm}
\begin{table}\label{tabZ}
\tabnum: eigenvectors with $P^{2}/2\pi\in\Z$.

$$
\begin{array}{|c|c|c|}\hline
P^{2}/2\pi=L_{0}^{(2)}&S=S_{3}=A_{3}^{(2)}&\ket{\{n_{k}\}}\\\hline
0&0&\ket{0}\\ 
2&0&\ket{1_{1}1_{-1}}\\ 
4&0&\ket{2_{1}2_{-1}}\\
1&1&\ket{1_{1}}\\ 
3&1&\ket{1_{3}}\\\hline
\end{array}
$$

\end{table}

\begin{lemma}\label{LZalln} (analog of PII-L9):\rm
~the states from Table~\ref{tabZ} are annulated by
$\alpha_{2}^{+}$ and have integer-valued 
$P^{2}/2\pi$.
\end{lemma}

\begin{lemma}\label{LZn1} (leading term):\rm
~let's keep in $\alpha_{2}^{+}$ only 
the leading $(1/\tilde n_{1})$-term:
$\alpha_{2}^{+}|_{n=1}=(-\Sigma_{-}+S_{-}/\gamma)/\tilde n_{1}$. 
The states from the first two Regge-trajectories
are annulated by $\alpha_{2}^{+}|_{n=1}$. In this approximation
the first two Regge-trajectories have integer-valued $P^{2}/2\pi$
and are linear: $P^{2}/2\pi=S+k$, $k=0,2$.
\end{lemma}

\begin{lemma}\label{LZlimc} (heavy vacuum modes): \rm
~in the limit $1\ll c_{1}^{2}\ll c_{2}\ll c_{1}^{4}$ 
the spectrum of $(P^{2}/2\pi,S)$ tends to the spectrum of
$(L_{0}^{(2)},A_{3}^{(2)}\geq0)$.
\end{lemma}


\noindent{\it Remark} (other ordering rules). We have fixed the ordering 
$\alpha_{2}\alpha_{2}^{+}$ in the definition of mass-square operator
after testing a number of alternative rules:
$a_{2}^{+}a_{2}$, $\alpha_{2}^{+}\alpha_{2}$, etc.
The alternative rules produce less regular spectrum $(P^{2}/2\pi,S)$.
Particularly, for the ordering $\alpha_{2}^{+}\alpha_{2}$
the operator $S_{+}$, present in $\alpha_{2}\sim(-\Sigma_{+}+S_{+}/\gamma)$,
does not annulate the states on the leading Regge trajectory
$S_{3}=S=P^{2}/2\pi$, and introduces additional non-linearity 
to this trajectory.
For the ordering $a_{2}^{+}a_{2}$ the operator $a_{1}$,
standing on the right in $a_{2}$, annulates the vacuum state
at each $S$-level: $a_{2}\ket{L_{0}^{(2)}=0,S}=0$,
therefore $P^{2}\ket{L_{0}^{(2)}=0,S}=0$ for any $S$,
the resulting spectrum looses Regge behaviour.
In other tested definition we performed a search of the pairs
$a_{1}^{*}a_{1}$ in $P_{n}$ on the classical level, and
later replace them $a_{1}^{+}a_{1}\to\tilde n_{1}$. This definition
creates almost linear Regge trajectories for small $S$,
possessing more non-linearity at large $S$.

\section*{Conclusion}

In this series of papers we have demonstrated a possibility
of construction of the quantum theory of Nambu-Goto string
in the space-time of dimension $d=4$. The general approach is
the selection of the light-cone gauge with the gauge axis
related in Lorentz-invariant way with the world sheet.
In this approach the Lorentz group transforms the world sheet 
together with the gauge axis and is not 
followed by reparametrizations. As a result, the theory
becomes free of anomalies in Lorentz group and in the group
of internal symmetries of the system. The constructed
quantum theory possesses spin-mass spectrum with Regge-like
behaviour.

Certain problems are still present in this theory,
which however do not hinder its implementations, 
e.g. for the construction of string models of the hadrons.
The results, produced by the theory, are influenced
by ordering of operators and other details of quantization procedure.
The theory does not contain {\it algebraic anomalies}, but possesses 
the features, which can be called {\it spectral anomalies}.
Particularly, Hamiltonian $P^{2}/2\pi$, classically generating
$2\pi$-periodic evolution, in quantum theory is influenced
by ordering rules and does not have strictly equidistant spectrum. 
This fact does not create problems for the hadronic models,
where this spectrum is subjected to phenomenological corrections
and experimentally is not strictly equidistant as well. 
The theory also 
possesses a topological defect, appearing as a discrete gauge
symmetry, identifying the points in the phase space (Gribov's copies).
This classical symmetry is related with discrete non-linear 
reparametrizations of the world sheet and is not preserved 
on the quantum level. In our construction we use the expansion series
in the vicinity of one Gribov's copy, by these means distinguishing it
in the quantum theory. We have also shown that
the leading term of the expansion, which has a minimal
ordering ambiguity and is easier for computation, is sufficient 
to reproduce Regge behaviour of the spectrum. Therefore,
practically one can keep this term to describe the main effect
and include further terms in the form of phenomenological corrections,
together with the contributions of other nature 
[16-21]:
gluonic tube thickness, quark masses, charges, 
spin-orbital interaction, etc.


\def\vsp{\vspace{-2mm}}

\vsp\paragraph*{Appendix 1:} proofs for the lemmas.

\vspace{2mm}\noindent \ref{Lgrib1}. 
In the space of variables 
$M=(\vec e_{3},{\vec q}_{n\perp},{\vec p}_{n\perp})$,
introduced in Part~II, the subgroup generated by $\chi_{3}$
acts trivially, and the whole gauge group of $\chi_{i}$-constraints
becomes equivalent to $SO(3)/SO(2)\sim S^{2}$. This group defines
the vector fields ${\vec q}_{n\perp}(\vec e_{3}),
{\vec p}_{n\perp}(\vec e_{3})$, tangent to the sphere $|\vec e_{3}|=1$. 
Each of these vector fields defines an embedding 
$f:~S^{2}\to TM(S^{2})$ of the sphere to its tangent bundle. Let $\tilde G$
be an image of this embedding for the vector field ${\vec q}_{s\perp}$.
It is a smooth 2-dimensional submanifold of 4-dimensional tangent bundle.
The gauge condition ${\vec q}_{s\perp}=0$
defines another 2-dimensional submanifold $\tilde F$,
which can be identified with original sphere $S^{2}$.
Due to a theorem \cite{DNF} p.525, the index of intersection 
for these manifolds coincides with the index of singular points 
for the corresponding vector field. Then we trivially lift this 
construction to the whole space $M$.


\vspace{2mm}\noindent \ref{Lgrib_ph}. 
The orbits of Gribov's copies are defined as zero level
curves of the function $F(\vec e_{3})=
(\vec q_{n}\times \vec p_{n},\vec e_{3})\sim 
i(\vec a_{n}\times \vec a_{n}^{*},\vec e_{3})$. This function does not
depend on phase rotations, thus the orbits are not changed
after the multiplication by the phase factor.
It's easy to verify that in both definitions the evolution 
$a_{n}\to a_{n}e^{-in\tau}$ is equivalent to the motion of the marked
point $O$ along the supporting curve $\vec Q(\s)$ under the action of 
uniform shifts: $\s\to\s+\tau$. The only difference is that
for PII-Eq.(12) these shifts are performed 
in lcg-parametrization, while (\ref{oint2}) corresponds 
to the natural parametrization.

\vspace{2mm}\noindent \ref{Lgrib2}. The structure of singularities
near the poles has been described in PII-L5,6. 
On the whole sphere of $\vec e_{3}$ 
for the straight string solution $\vec Q(\s)=(\cos\s,\sin\s,0)$ 
the integral (\ref{oint2}) can be evaluated in terms of Bessel functions:
$\vec a_{n}=(-i)^{n}\sqrt{\pi/8}( 
J_{n+1}(n\rho)e^{i(n+1)\alpha}+J_{n-1}(n\rho)e^{i(n-1)\alpha}, $ $
-iJ_{n+1}(n\rho)e^{i(n+1)\alpha}+iJ_{n-1}(n\rho)e^{i(n-1)\alpha},0)$, 
where $\vec e_{3}=(\rho\cos\alpha,\rho\sin\alpha,\rho_{z})$, 
$0\leq\rho\leq1$, $\rho^{2}+\rho_{z}^{2}=1$, $n=1,2...$ The real part 
$\vec q_{n}=\mbox{Re}(\vec a_{n})=\sqrt{\pi/8}(J_{n+1}(n\rho)\cos\phi_{+}
+J_{n-1}(n\rho)\cos\phi_{-},J_{n+1}(n\rho)\sin\phi_{+}
-J_{n-1}(n\rho)\sin\phi_{-},0)$, where $\phi_{\pm}=(n\pm1)\alpha-\pi n/2$.
Zeros of $\vec q_{n\perp}=\vec q_{n}-(\vec q_{n}\vec e_{3})\vec e_{3}$
correspond to $\vec q_{n}\times\vec e_{3}=(q_{n}^{2}e_{3}^{3},
-q_{n}^{1}e_{3}^{3},q_{n}^{1}e_{3}^{2}-q_{n}^{2}e_{3}^{1})=0$.
The solutions disjoin to the following branches:

(b1) $e_{3}^{3}=0$, equator: $q_{n}^{1}e_{3}^{2}-q_{n}^{2}e_{3}^{1}=0$;

(b2) $e_{3}^{3}\neq0$, $q_{n}^{1}=q_{n}^{2}=0$.

For the case (b1) $\rho=1$ and after simplifications we have
$(J_{n-1}(n)-J_{n+1}(n))\cdot\sin(n\alpha-\pi n/2)=0$,
i.e. $\alpha=\pi/2+\pi k/n$, $k=0...2n-1$.
Considering the expansion of the vector field e.g. near the point
$\alpha=\pi/2$ in the local coordinates $(\Delta\alpha,\rho_{z})$,
we have $q_{n\perp}\sim(-n(J_{n-1}(n)-J_{n+1}(n))\Delta\alpha,
-(J_{n-1}(n)+J_{n+1}(n))\rho_{z})$. Because of the property
$J_{n-1}(n)>J_{n+1}(n)>0$ (see below) we conclude that 
this singular point is the node.
For the vector fields, corresponding to the straight line string, 
the evolution is equivalent to the rotation about z-axis. 
The evolution during $\Delta\tau_{n}=\pi/n$ matches the node 
on the equator to the neighbour one. Let's consider small perturbations
of the string from the straight configuration. Non-degenerate points
persist and preserve their type after the deformation. From PII-L3
we know that singular points move along zero-level curves of the
function $F(\vec e_{3})=(\vec q_{n}\times \vec p_{n},\vec e_{3})$.
For the straight line string this function is
$$F\sim i{\footnotesize
\left|\begin{array}{ccc}\rho\cos\alpha&\rho\sin\alpha&\rho_{z}\\
a_{n}^{1}&a_{n}^{2}&0\\ a_{n}^{1*}&a_{n}^{2*}&0
\end{array}\right|}\sim \rho_{z}(J_{n-1}^{2}(n\rho)-J_{n+1}^{2}(n\rho)).$$
\noindent
Near the equator $F\sim\rho_{z}$, i.e. the equator is non-degenerate
zero level curve of this function: $\df F/\df\rho_{z}\neq0$. 
Parameterizing zero level curve by the longitude: $\vec e_{3}(\alpha)$,
after small deformations: $F(\vec e_{3}+\delta\vec e_{3})+
\delta F(\vec e_{3})=0$, for the shifts of the latitude 
we have: $\delta\rho_{z}=-(\df F/\df\rho_{z})^{-1}\delta F$.
Therefore, the curve will be slightly deformed in the
vicinity of the equator. It was proven in the Part~II,
that the vector field reverses its sign after the lapse of time
$\Delta\tau_{n}$. During this time the pattern of singularities
returns to the initial state. Thus, the nodes near the equator should 
move to the neighbor ones, to satisfy the continuous limit 
with the straight string configuration.

For the case (b2) $\rho<1$ and we have a system 
$$\left\{\begin{array}{l}
J_{n-1}(n\rho)\sin(\phi_{+}+\phi_{-})=0,\\
J_{n+1}(n\rho)+J_{n-1}(n\rho)\cos(\phi_{+}+\phi_{-})=0.
\end{array}\right. $$
\noindent Here we again have two branches:

(b21) $J_{n+1}(n\rho)=J_{n-1}(n\rho)=0$;

(b22) $\phi_{+}+\phi_{-}=\pi k$, $J_{n+1}(n\rho)+(-)^{k}J_{n-1}(n\rho)=0$.

\begin{figure}\label{fcwetok}
\begin{center}
~\epsfxsize=4cm\epsfysize=4cm\epsffile{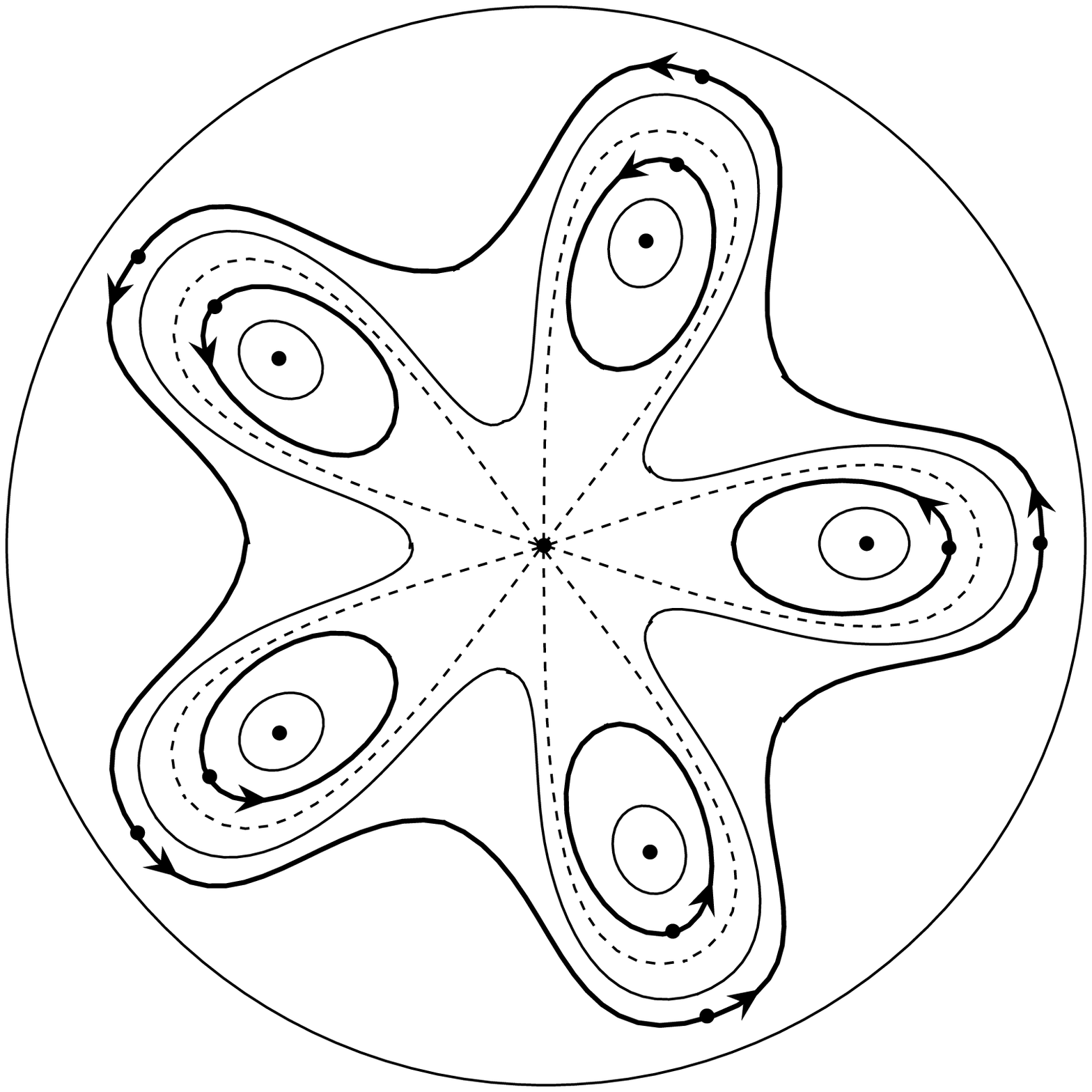}

\fignum Unfolding of multi-saddle, $n=6$.
\end{center}
\end{figure}

In the case (b21) we have only solution $\rho=0$ at $n>1$.
It gives pole singularities described in PII-L5,6.
Near this solution $J_{n+1}\ll J_{n-1}\ll1$
and in the passage around this point
the vector $(q_{n}^{1},q_{n}^{2})\sim \rho^{n-1}
(\cos\phi_{-},-\sin\phi_{-})$ performs $(n-1)$ revolutions
in the direction opposite to the passage. It corresponds to
the saddle for $n=2$ and multi-saddle for $n>2$.
In the complex plane of $z=\rho e^{i\alpha}$ the mapping
$z\to q=q_{n}^{1}+iq_{n}^{2}$ near this point
can be represented as $q\sim(z^{*})^{n-1}$.
Considering small deformations of the string 
from straight configuration, we introduce a correction
$(z^{*})^{n-1}+\delta q$, where $\delta q(z,z^{*})=\delta q_{0}+
\delta q_{0z}z+\delta q_{0z^{*}}z^{*}+...$ is small 
together with its derivatives, and in small vicinity of
the singular point we keep only the constant term $\delta q_{0}$.
At a given point there is 1-1 correspondence between the set
$(q_{n},p_{n})_{0}$ and the shape of supporting curve,
and the arbitrary small variations $\delta q_{0}$
can be reproduced by variations of supporting curve.
That's why we can apply the same perturbation analysis 
as for general vector fields, in spite of the fact that
we consider the fields of special form.
New singular points are $z_{k}=(-\delta q_{0}^{*})^{1/(n-1)}e^{2\pi ik/(n-1)}$,
$k=0...n-2$. Expansion of the field in their vicinity has a form
$q=(z_{k}^{*}+\Delta z^{*})^{n-1}+\delta q_{0}\sim
(n-1)(z_{k}^{*})^{n-2}\Delta z^{*}$, therefore we have
$(n-1)$ saddles. Considering their evolution:
$q\to q\cos n\tau+p\sin n\tau$, where $p\sim i(z^{*})^{n-1}+\delta p$,
we have $(z^{*})^{n-1}e^{in\tau}+\delta q\cos n\tau+\delta p\sin n\tau=0$,
i.e. $z_{k}=(-\delta z_{1}^{*}-\delta z_{2}^{*}e^{2in\tau})^{1/(n-1)}
e^{2\pi ik/(n-1)}$, where $\delta z_{1}=(\delta q-i\delta p)/2$, 
$\delta z_{2}=(\delta q+i\delta p)/2$. 
If $n=2$, we have a single saddle moving along a closed loop with a period
$\Delta\tau_{2}=\pi/2$, as described in PII-L6. 
If $n>2$, the dynamics is more complex, see \fref{fcwetok}.
Particularly, if $\delta z_{1}=0$, $\delta z_{2}\neq0$, $n-1$ saddles
move along a common circle, matching the neighbour one after 
the lapse of time $\Delta\tau_{n}=\pi/n$. 
If $\delta z_{1}\neq0$, $\delta z_{2}=0$, the saddles stay fixed. 
Generally, if $0<|\delta z_{2}|<|\delta z_{1}|\ll1$,
the origin $z=0$ is placed outside the circle 
$\delta z_{1}+\delta z_{2}e^{-2in\tau}$, 
and the complex root of $(n-1)$-th order
gives $(n-1)$ saddles moving along disjoint loops 
with a period $\Delta\tau_{n}$.
If $0<|\delta z_{1}|<|\delta z_{2}|\ll1$, the origin is inside the circle 
and the saddles move along a common loop.
The separatriss $|\delta z_{1}|=|\delta z_{2}|$
for $n>3$ is structurally unstable and under the influence of 
higher order corrections is unfolded to further complex cases.

Note: this phenomenon resembles the bifurcation 
of degenerate critical point of the function
$f=x^{3}-3xy^{2}$ \cite{Arnold-lectures} (known also
as ``monkey saddle'' \cite{Gray-monkey-saddle}),
and the structure of phase flows near high order resonanses 
of Hamiltonian systems \cite{Arnold-mat-phys}.

In the case (b22) we have $\alpha=\pi k/2n$, $k=0...4n-1$,
$J_{n+1}(n\rho)-J_{n-1}(n\rho)=0$ for odd $k$
and $J_{n+1}(n\rho)+J_{n-1}(n\rho)=0$ for even $k$.
The solution $\rho=0$, $n>1$ has been already considered.
There are no solutions in the interval $0<\rho\leq1$
due to the inequality $J_{n-1}(n\rho)>J_{n+1}(n\rho)>0$,
which follows from easily provable relation 
for Bessel functions: $J_{n-1}(x)>J_{n}(x)>J_{n+1}(x)>...>0$
for $0<x\leq n$, $n=1,2,3...$

\vspace{2mm}\noindent \ref{Lstruct}. 
The equations (6) of Part II in the arbitrary spin case
can be rewritten as
\begin{eqnarray}
&&a_{2}={a_{1}^{2}\over n_{1}^{2}}\cdot \left(
-z+{S_{+}\over\gamma}(1+\beta)\right),\label{a2bet}\\
&&a_{2}^{*}={a_{1}^{*2}\over n_{1}^{2}}\cdot \left(
-z^{*}+{S_{-}\over\gamma}(1+\beta)\right),\nn\\
&&\beta=a_{2}^{*}a_{2}\gamma^{2}-\beta^{2}/2,\nn
\end{eqnarray}
where $n_{1}=a_{1}^{*}a_{1}$, 
$z=\Sigma_{+}+f^{*}a_{2}^{*}+{g'}^{*}a_{2}+d^{*}a_{2}^{*2}+
d a_{2}a_{2}^{*}/2$, $(1+\beta)/\gamma=\sqrt{P^{2}/2\pi}$.
Substituting the expansions 
$$a_{2}=\sum_{n\geq1}{P_{n}a_{1}^{2}\over n_{1}^{2n}},\quad
\beta=\sum_{n\geq1}{b_{n}\over n_{1}^{2n}},$$
we obtain the following recurrent relations:

\baselineskip=0.4\normalbaselineskip\footnotesize

\begin{eqnarray}
&&~\hspace{-8mm}
P_{1}=-\Sigma_{+}+{S_{+}\over\gamma},\quad b_{1}=0,\nn\\
&&~\hspace{-8mm}
P_{n}=-f^{*}a_{1}^{*2}P_{n-1}^{*}-{g'}^{*}a_{1}^{2}P_{n-1}
-d^{*}a_{1}^{*4}\sum_{1\leq m\leq n-2}P_{n-m-1}^{*}P_{m}^{*}\nn\\
&&~\hspace{-8mm}
-\half d a_{1}^{*2}a_{1}^{2}\sum_{1\leq m\leq n-2}P_{n-m-1}P_{m}^{*}
+{S_{+}\over\gamma}b_{n-1},\ n>1,\nn\\
&&~\hspace{-8mm}
b_{n}=a_{1}^{*2}a_{1}^{2}\gamma^{2}\sum_{1\leq m\leq n-1}
P_{n-m}^{*}P_{m}
-\half\sum_{1\leq m\leq n-1}b_{n-m}b_{m},\ n>1.\nn
\end{eqnarray}

\baselineskip=\normalbaselineskip\normalsize

Then the properties (i)-(v) can be easily proven  
by induction. The property (i) also follows from the scaling symmetry 
of the system (\ref{a2bet}): formal transformations
$\{ n_{1}\to\sqrt{C}\cdot n_{1}$, $v\to C\cdot v$, 
$v\in(d,f,g',\Sigma_{-},S_{-}$ and their conjugates$)\}$
and $\{n_{1}\to C\cdot n_{1}$, $a_{1}\to C\cdot a_{1}\}$
preserve $a_{2},\beta$; the property (iii) follows 
from reflection symmetry: $a_{1}\to -a_{1}$ preserves $a_{2},\beta$.
The property (ii) follows from the fact that 
$a_{2}=\beta=0$ is the solution of (\ref{a2bet}) at $\Sigma_{\pm}=S_{\pm}$.

\begin{figure}\label{fLconv}
\begin{center}
~\epsfxsize=8cm\epsfysize=3cm\epsffile{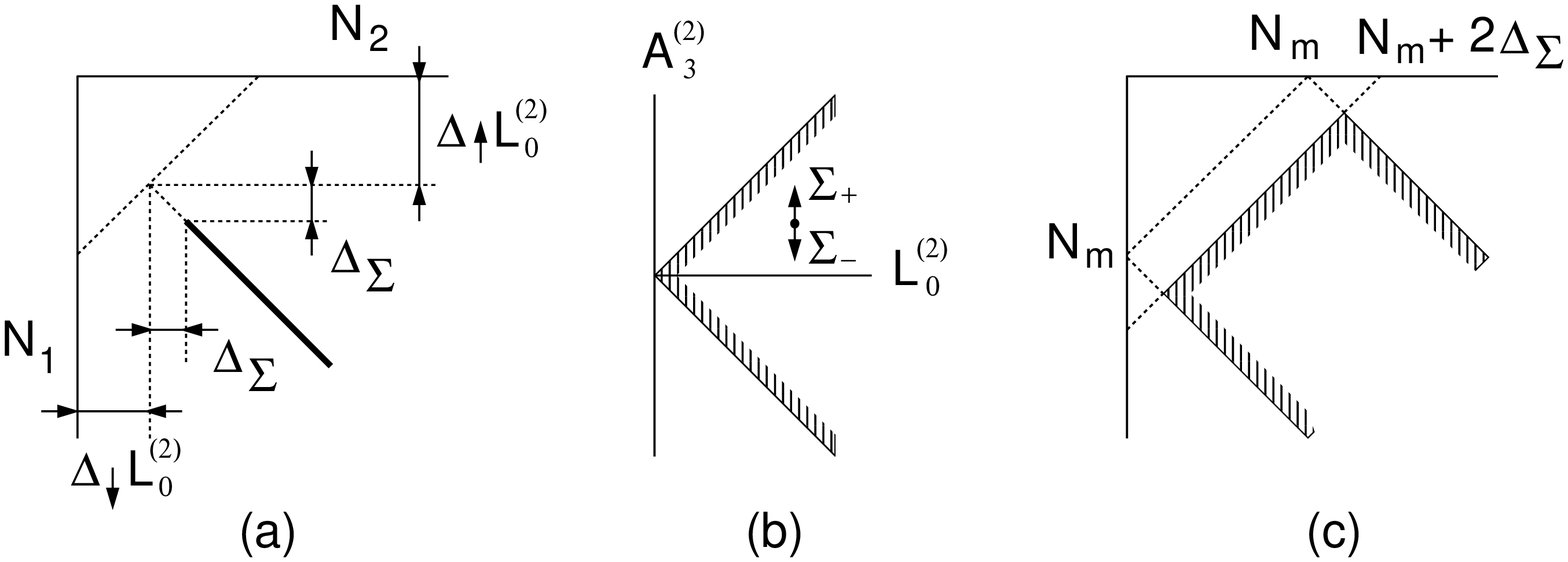}

\fignum To the proof of lemma \ref{Lconv}.
\end{center}
\end{figure}

\vspace{2mm}\noindent \ref{Lconv}. 
Let's consider a monomial of $P_{n}$,
substitute the definitions of elementary operators 
$d=d_{+}-d_{-},\ f=f_{+}-f_{-},\ g'=g'_{-}-g_{+}$ 
(see PII-Eq.(7)), expand it to a number of (secondary) monomials 
and normally order each one. 
Let $n(d_{\uparrow}),n(f_{\uparrow}),n(g_{\uparrow}),n(a_{1\uparrow})$ 
be the numbers of $L_{0}^{(2)}$-raising operators in the secondary monomial,
$n(d_{\downarrow}),n(f_{\downarrow}),n(g_{\downarrow}),n(a_{1\downarrow})$ 
be the numbers of $L_{0}^{(2)}$-lowering operators. 
Using the definitions of the charges (Table~1 in Part II), we see that 
the secondary monomial is represented as a diagonal with offsets
$\Delta_{\uparrow}L_{0}^{(2)}=\mbox{$4n(d_{\uparrow})$}
+\mbox{$2n(f_{\uparrow})$}+\mbox{$2n(g'_{\uparrow})$}
+n(a_{1\uparrow})$,
$\Delta_{\downarrow}L_{0}^{(2)}=
\mbox{$4n(d_{\downarrow})$}+\mbox{$2n(f_{\downarrow})$}+
\mbox{$2n(g'_{\downarrow})$}+n(a_{1\downarrow})$ in the matrix,
displayed at \fref{fLconv}a:
$\bra{L_{0}^{(2)}=N_{1}} mon \ket{L_{0}^{(2)}=N_{2}}\sim
\bra{L_{0}^{(2)}=N_{1}-\Delta_{\uparrow}L_{0}^{(2)}} 
diag \ket{L_{0}^{(2)}=N_{2}-\Delta_{\downarrow}L_{0}^{(2)}}.$
Here $diag$ represents $L_{0}^{(2)}$-neutral part of the monomial,
consisting of the operators $\Sigma_{\pm},S_{\pm},\gamma$,
and non-zero value of this matrix element corresponds to 
$N_{1}-\Delta_{\uparrow}L_{0}^{(2)}=N_{2}-\Delta_{\downarrow}L_{0}^{(2)}\geq0$.

Then, using the inequality $|A_{3}^{(2)}|\leq L_{0}^{(2)}$
and charge properties of $\Sigma_{\pm}$, see \fref{fLconv}b,
we conclude that non-zero value for the considered matrix element 
corresponds to the following restrictions on the number of 
$\Sigma_{\pm}$-operators: 
$n(\Sigma_{+})\leq2(N_{2}-\Delta_{\downarrow}L_{0}^{(2)})$, 
$n(\Sigma_{-})\leq2(N_{2}-\Delta_{\downarrow}L_{0}^{(2)})$,
i.e. $N_{2}\geq\Delta_{\downarrow}L_{0}^{(2)}+\Delta_{\Sigma}$,
$\Delta_{\Sigma}=\max(n(\Sigma_{+}),n(\Sigma_{-}))/2$.
As a result,  non-zero entries in the diagonal, shown on \fref{fLconv}a,
receive an additional offset. After summation over all secondary monomials
non-zero entries of the matrix are localized
in the region, shown on \fref{fLconv}c. Here 
$N_{m}=\Delta_{\uparrow}L_{0}^{(2)}+\Delta_{\downarrow}L_{0}^{(2)}
=4n(d)+2n(f)+2n(g)+n(a_{1})$ and $n(x)$ represents the number
of variables $(x,x^{*})$ in the primary monomial. 
Particularly, non-zero entries are localized at
$N_{1}+N_{2}\geq N_{m}+2\Delta_{\Sigma}$.
For the number of $S_{\pm}$-operators we have the following restrictions:
$n(S_{+})\leq2S$, $n(S_{-})\leq2S$. Then, using the property (i)
from the structural lemma: 
$n(S_{\pm},\Sigma_{\pm},d,f,g)=n$, $n(a_{1})=2(n-1)$, 
we construct the following system of 11 linear
inequalities and 1 linear equation:

\baselineskip=0.4\normalbaselineskip\footnotesize

\vspace{-2mm}
\begin{eqnarray}
&&\hspace{-5mm}
4n(d)+2n(f)+2n(g)+n(\Sigma_{+})\leq N_{1}+N_{2}+2-2n,\nn\\
&&\hspace{-5mm}
4n(d)+2n(f)+2n(g)+n(\Sigma_{-})\leq N_{1}+N_{2}+2-2n,\nn\\
&&\hspace{-5mm}
0\leq n(S_{+})\leq2S,\quad 0\leq n(S_{-})\leq2S,\nn\\ 
&&\hspace{-5mm}
n(\Sigma_{+})\geq0,\ n(\Sigma_{-})\geq0,\ 
n(d)\geq0,\ n(f)\geq0,\ n(g)\geq0,\nn\\
&&\hspace{-5mm}
n(d)+n(f)+n(g)+n(\Sigma_{+})+n(\Sigma_{-})+n(S_{+})+n(S_{-})=n.\nn
\end{eqnarray}

\baselineskip=\normalbaselineskip\normalsize

The inequalities define a convex polyhedron in 9-dimensional space
$(n,N_{1}+N_{2}+2,S,n(d),n(f)+n(g),n(\Sigma_{+}),n(\Sigma_{-}),
n(S_{+}),n(S_{-}))$. The linear equation defines its 8-dimensional slice. 
Projecting it to a 3-dimensional subspace of parameters 
$(n,N_{1}+N_{2}+2,S)$, we obtain a convex polyhedron,
which represents the region of existence of the solutions.
Solving the system by {\it Mathematica},
we have $n\leq\min\{1+(N_{1}+N_{2})/2,\ (4+2(N_{1}+N_{2})+4S)/5\}$.
This condition defines a region, where non-zero entries of 
the matrix element considered in the lemma are located.


\vspace{2mm}\noindent \ref{LZalln}. 
The states, annulated by $\alpha_{2}^{+}$, have
$P^{2}/2\pi=L_{0}^{(2)}\in\Z$, at $c_{3}=0$.
Due to the property (ii) of structural lemma
each monomial in $P_{n}$ contains at least one operator from the group
$(\Sigma_{\pm},S_{\pm})$. After the normal ordering some number of
$L_{0}^{(2)}$-lowering operators can also appear on the right of 
$\Sigma_{\pm}$. Due to the charge properties 
and the property (iii) of lemma \ref{Lstruct}, the $L_{0}^{(2)}$-lowering operators 
together decrease $L_{0}^{(2)}\to L_{0}^{(2)}-2k$, $k\in\Z,\ k\geq1$.
The vacuum state $(P^{2}/2\pi,S)=(L_{0}^{(2)},A_{3}^{(2)})=(0,0)$
is annulated by all these operators. The state 
$(P^{2}/2\pi,S)=(L_{0}^{(2)},A_{3}^{(2)})=(1,1)$ is annulated by
$\Sigma_{\pm}$-operators, together with other states on the leading Regge 
trajectory, see \fref{elka-corrected}. If there are any $L_{0}^{(2)}$-lowering 
operators at the right, they decrease $L_{0}^{(2)}$ by $\geq2$ and
annulate this state. It is also annulated by $S_{-}$ and generally
is not annulated by $S_{+}$. However, due to the property (iv)
in lemma \ref{Lstruct} each $S_{+}$ enters to $\alpha_{2}^{+}$ together with
$a_{1}^{2}$, which annulates this state. For the state 
$(P^{2}/2\pi,S)=(L_{0}^{(2)},A_{3}^{(2)})=(2,0)$ the terms with $S_{\pm}$
are inactive. $\Sigma_{\pm}$ annulate this state, see \fref{elka-corrected}.
If there are any $L_{0}^{(2)}$-lowering operators at the right, they 
can transform this state only to vacuum, and the vacuum is also
annulated by $\Sigma_{\pm}$. Similarly, on the states
$(P^{2}/2\pi,S)=(L_{0}^{(2)},A_{3}^{(2)})=(4,0)$ 
the terms with $S_{\pm}$ are inactive. There are 3 states 
on the level $(L_{0}^{(2)},A_{3}^{(2)})=(4,0)$,
1 state on the level $(L_{0}^{(2)},A_{3}^{(2)})=(4,1)$ and 
1 state on the level $(L_{0}^{(2)},A_{3}^{(2)})=(4,-1)$, 
see \fref{elka-corrected}.
As a result, the operator $\Sigma_{+}$ annulates 2-dimensional subspace
of $(L_{0}^{(2)},A_{3}^{(2)})=(4,0)$ and one state in this subspace is
also annulated by $\Sigma_{-}$. Using the definition of $\Sigma_{\pm}$, 
it's easy to verify that $\Sigma_{\pm}\ket{2_{1}2_{-1}}=0$. 
If there are any $L_{0}^{(2)}$-lowering operators at the right 
of $\Sigma_{\pm}$, 
they can transform this state to the states with $L_{0}^{(2)}=0,2$, 
which are also annulated by $\Sigma_{\pm}$, see \fref{elka-corrected}. 
There are 2 states on the level $(L_{0}^{(2)},A_{3}^{(2)})=(3,1)$:
$\ket{1_{3}},\ket{2_{1}1_{-1}}$. Both states are annulated by 
$\Sigma_{\pm}$. If there are any $L_{0}^{(2)}$-lowering operators 
at the right, 
they transform these states to $L_{0}^{(2)}=1$, also annulated by 
$\Sigma_{\pm}$. Both states are annulated by $S_{-}$. Generally
they are not annulated by $S_{+}$, and due to the property (iv)
in lemma \ref{Lstruct} each $S_{+}$ enters to $\alpha_{2}^{+}$ together with
$a_{1}^{2}$, which annulates one of these states: $\ket{1_{3}}$.

\vspace{2mm}\noindent \ref{LZn1}. 
In $\alpha_{2}^{+}|_{n=1}=(-\Sigma_{-}+S_{-}/\gamma)/\tilde n_{1}$
the operator $\Sigma_{-}$ decreases $A_{3}^{(2)}$ by 1,
the operator $S_{-}$ increases $S_{3}$ by 1.
Let's consider the states $A_{3}^{(2)}=S_{3}=S\geq0$,
annulated by $S_{-}$. On \fref{elka-corrected} we see
that $\Sigma_{-}$ annulates the leading Regge trajectory,
because there are no states immediately below this trajectory.
Concerning the second trajectory $0\leq A_{3}^{(2)}=L_{0}^{(2)}-2$,
one state at $(L_{0}^{(2)},A_{3}^{(2)})=(2,0)$ and two states
at $(L_{0}^{(2)},A_{3}^{(2)})=(3,1)$ are annulated by $\Sigma_{-}$
due to the same reason, while the sequence of double states
$2\leq A_{3}^{(2)}=L_{0}^{(2)}-2$ is transformed by the operator 
$\Sigma_{-}$ to the sequence of non-degenerate states 
$1\leq A_{3}^{(2)}=L_{0}^{(2)}-3$, thus annulating one state
from each pair. As a result, the described states are annulated
by $\alpha_{2}^{+}|_{n=1}$ and in the considering approximation
these states have $P^{2}/2\pi=L_{0}^{(2)}$, 
from here we obtain the statement of the lemma.

\vspace{2mm}\noindent \ref{LZlimc}. 
$\alpha_{2}^{+}=\sum P_{n}^{+}/\tilde n_{1}^{2n-1}$, 
where $P_{n}=O(\gamma^{-[(n+1)/2]})$ 
due to the property (v) of lemma \ref{Lstruct}.
Using the definitions $\tilde n_{1}=a_{1}^{+}a_{1}+c_{1}$,
$\gamma=(L_{0}^{(2)}+c_{2})^{-1/2}$, 
in the limit $1\ll c_{1}^{2}\ll c_{2}\ll c_{1}^{4}$ we have
$\alpha_{2}^{+}|_{n}\sim c_{2}^{[(n+1)/2]/2}/c_{1}^{2n-1}$,
i.e. $\alpha_{2}^{+}|_{n=1}\sim c_{2}^{1/2}/c_{1}\gg 1$,
$\alpha_{2}^{+}|_{n=2}\sim c_{2}^{1/2}/c_{1}^{3}\ll 1/c_{1}\ll 1$,
$\alpha_{2}^{+}|_{n\geq3}\sim c_{2}^{[(n+1)/2]/2}/c_{1}^{2n-1}
\leq c_{2}^{(n+1)/4}/c_{1}^{2n-1}\ll 1/c_{1}^{n-2}\ll 1$.
Writing explicitly the leading contribution:
$\alpha_{2}^{+}=S_{-}\cdot c_{2}^{1/2}/c_{1}+o(1)$,
we see that the states with $S=S_{3}=A_{3}^{(2)}$
are annulated by $\alpha_{2}^{+}$ in the described approximation,
while other states receive large contribution to  $P^{2}/2\pi$
and are shifted away from the origin of the spectrum.
Due to this property, the spectrum of $(P^{2}/2\pi,S)$ 
tends to the spectrum of $(L_{0}^{(2)},A_{3}^{(2)}\geq0)$,
provided that $c_{3}=0$.

\vsp\paragraph*{Appendix 2:} super-selection rule.

In several earlier works the quantum top appeared 
in the context of relativistic string dynamics. 
Particularly, the paper \cite{3str} describes a special family 
of motions in a theory of Y-shaped strings, globally isomorphic 
to the phase space of the top $\R^{3}\times SO(3)$.
In the paper \cite{2par} a similar family 
was selected in the phase space of open strings. 
In the present paper we also have $SO(3)$-factor 
in the topology of the phase space. As a manifold,
it has a fundamental group $\pi_{1}(SO(3))=\Z_{2}$,
i.e. it possesses two homotopically non-equivalent classes 
of contours. The contour, representing one complete
revolution, cannot be contracted to a point (unit of the group),
while two revolutions are contractible \cite{Casimir}. This fact gives
a possibility to construct ray representations
of rotation group, corresponding to integer and half-integer 
values of spin \cite{Weyl}. The existence of half-integer spin 
and associated double-valued representations of $SO(3)$
is also explained in terms of path integration technique:
non-homotopic contours can contribute to path integrals 
with different phase factors \cite{3str_ref9,3str}. 

The question on the availability of half-integer spin 
is subtle and requires special consideration 
in each particular problem. Now we will consider a model example, 
representing the main features of our problem:
the mechanics of 3 particles in 3-dimensional space, 
$(\vec x_{n},\vec p_{n})\in\R^{9}\times\R^{9}$. 
Let's apply the following canonical transformations:
$\Omega=d\vec p_{n}\wedge
d\vec x_{n}=d\vec P\wedge d\vec X+\half d(\vec S\times\vec e_{i})
\wedge d\vec e_{i}+d\rho_{ni}\wedge dq_{ni},$
where $\vec X=\sum m_{n}\vec x_{n}/\sum m_{n}$,
$m_{n}$ are masses of the particles,
$\vec P=\sum \vec p_{n}$, $\vec q_{n}=\vec x_{n}-\vec X$,
$\vec\rho_{n}=\vec p_{n}-m_{n}/m_{3}\cdot\vec p_{3}$,
$\vec S=\sum \vec q_{n}\times\vec p_{n}=\sum \vec q_{n}\times\vec\rho_{n}$,
$\vec e_{i}$ is an orthonormal basis, $\rho_{ni}=\vec\rho_{n}\vec e_{i}$, 
$q_{ni}=\vec q_{n}\vec e_{i}$, and everywhere the sum over repeated 
indices is assumed. Let's fix the basis
with respect to the particles. The following cases should be
considered:

($s_{1}$) $\vec q_{1}$ and $\vec q_{2}$ 
are linearly independent. In this case we can select e.g. 
$\vec e_{1}\uparrow\uparrow\vec q_{1}$, 
$\vec e_{2}\uparrow\uparrow\vec q_{2\perp}=\vec q_{2}-
(\vec q_{2}\vec q_{1})/\vec q_{1}^{~2}\cdot \vec q_{1}$. 
The last term in the symplectic form can be written as 
$d\rho_{11}\wedge dq_{11}+d\rho_{21}\wedge dq_{21}+d\rho_{22}\wedge dq_{22}$.
Here $q_{11}>0$, $q_{22}>0$. The obtained phase space
has topology $(\R^{9})\times(\R^{6}\times SO(3))$,
where the first factor represents spin-momentum space,
the second factor is configuration space, $SO(3)$ 
is the configuration space of the top.

($s_{2}$) $\vec q_{1}\neq0$, $\vec q_{2}\neq0$, 
$\vec q_{1}\uparrow\uparrow\vec q_{2}$.
Let's direct $\vec e_{1}\uparrow\uparrow\vec q_{1}$.
The symplectic form can be written as $\Omega=
d\vec P\wedge d\vec X+d(\vec S\times\vec e_{1})\wedge d\vec e_{1}
+d\rho_{11}\wedge dq_{11}+d\rho_{21}\wedge dq_{21}$.
Here $\vec S\vec e_{1}=0$, $q_{11}>0$, $q_{21}>0$. 
The topology of the phase space is $\R^{10}\times TM(S^{2})$,
the configuration space is $\R^{5}\times S^{2}$.
Here the tangent bundle $TM(S^{2})$ represents 
the phase space of rotator.

($s_{3}$) $\vec q_{1}\neq0$, $\vec q_{2}\neq0$, 
$\vec q_{1}\uparrow\downarrow\vec q_{2}$.
Corresponds to the same mechanics as ($s_{2}$), but $q_{21}<0$.

($s_{4}$) $\vec q_{1}\neq0$, $\vec q_{2}=0$.
The same mechanics as ($s_{2}$), but $q_{21}=0$ and $\rho_{21}$ omitted:
phase space $\R^{8}\times TM(S^{2})$, configuration space 
$\R^{4}\times S^{2}$.

($s_{5}$) $\vec q_{1}=0$, $\vec q_{2}\neq0$.
The same mechanics as ($s_{4}$), but $q_{22}=0$ and $\rho_{22}$ omitted.

($s_{6}$) $\vec q_{1}=\vec q_{2}=0$. In this case $\vec S=0$
and $\Omega=d\vec P\wedge d\vec X$. The phase space is 
$\R^{3}\times\R^{3}$, scalar particle.

The archetypes (top, rotator, scalar) are essentially the same 
as described in Sec.2 of Ref.\cite{3str_ref9}. The important difference 
is that in our example they appear not as isolated cases, but 
as a stratification of the configuration space:
$\R^{6}(\vec q_{1},\vec q_{2})=\cup_{i=1}^{6}s_{i}$.
The strata $s_{2\mbox{-}6}$ have zero measure and $codim(s_{2,3})=2$,
$codim(s_{4,5})=3$, $codim(s_{6})=6$ in $\R^{6}$.
The exclusion of these zero measure subsets changes the
topological structure of configuration space.
Note that the original configuration space is 
simply connected and quantization of the problem in original
representation gives only integer values of spin.
All closed loops in the original configuration space  
are contractible. The contours in this space can 
be related by a homotopy, intersecting the strata $s_{2\mbox{-}6}$, 
as a result, all the contours should contribute to path integrals 
with the same phase factors. 
Therefore, only single-valued representation of $SO(3)$ 
can be used. We conclude that half-integer spin in the considered 
model example appears as an artefact of the representation and 
should be rejected.

In our problem constraint $S_{3}=A_{3}\in\Z$ selects 
only integer $S$, however, with formal modifications 
$A_{3}\to A_{3}+1/2$, classically vanishing after the recovery 
of Planck's constant, it's possible to introduce half-integer spin
to the theory. The above argumentation favors against it.
During our construction we reject zero measure subsets 
of singular cases (see Part~II). 
The original configuration space of open string theory is 
simply connected, all contours in it are homotopic
to each other. Therefore we should not introduce 
half-integer spin in this theory.

In the problems \cite{3str,2par} submanifold 
$\R^{3}\times SO(3)$, representing particular
type of string motion, was selected in the original phase space, 
representing all possible motions. The original
configuration space possessed trivial topology:
$x_{\mu}(\s)\in C^{\infty}[0,\pi]$ for open string
and $x_{\mu}^{i}(\s)\in C^{\infty}[0,\pi]$
with $x_{\mu}^{1}(\pi)=x_{\mu}^{2}(\pi)=x_{\mu}^{3}(\pi)$
for Y-shaped string. Therefore, half-integer spin representations
have right to exist in frames of the restricted models 
\cite{3str,2par}, but would disappear in full quantum theories.

\vsp\paragraph*{Appendix 3:} Gribov's copies in quantum mechanics.

The Gribov's copies of string theory can be separated
to those which belong to the same $\chi_{0}$-orbit and are transformed
one to the other by the quarter period of the evolution,
and those which belong to different orbits. The factorization
of the phase space with respect to $\chi_{0}$-transformation
identifies the copies on the same $\chi_{0}$-orbit.

The discrete symmetry $D_{4}=\exp[-i(\pi/2)L_{0}^{(2)}]$, 
mapping the Gribov's copies on the $\chi_{0}$-orbit 
one to the other, is associated with the quantum number 
$Q=L_{0}^{(2)}\mod4$. In the whole phase space,
containing the external variables $(X,P)$,
$\chi_{0}$-orbit is not closed, but is periodically repeated as
$X\to X+2Pn$, $n\in\Z$. Considering the evolution in this space,
the generator of discrete symmetry should be extended to
$D_{4}'=\exp[i(\pi/2)(P^{2}/2\pi-L_{0}^{(2)})]$.
Classically, for the Gribov's copies near the poles
this evolution coincides with $D_{4}''=\exp[i(\pi/2)\chi_{0}]$.
Indeed, the evolutions $D_{4}'$ and $D_{4}''$ 
connect the same Gribov's copies, but along 
different trajectories, particularly, $D_{4}'$ does not contain 
$S_{\pm}$-variables, influencing the direction of gauge axis, and 
$D_{4}''$ does. For the Gribov's copies near the equator the evolutions
$D_{4}'$ and $D_{4}''$ connect not the same, but neighbor copies. 
These formal differences become crucial in quantum mechanics,
where $L_{0}^{(2)}$ has integer-valued spectrum and 
$M^{2}/2\pi=L_{0}^{(2)}+2\alpha_{2}\alpha_{2}^{+}$ has not. 
Note that the number $Q$ is associated with $L_{0}^{(2)}$ 
and $D_{4}$-symmetry, defined in the phase space of 
internal variables, independent of $(X,P)$.

Let's write $\chi_{0}=(P^{2}-M^{2})/2\pi$,
where $M^{2}$ is a function of the internal variables and 
has a discrete spectrum. Let's consider the Hamiltonians
of interaction $H_{\mbox{\small int}}$, which generate transitions 
in this spectrum. Generally $H_{\mbox{\small int}}$ should not commute
with $\chi_{0}$: this would mean that $H_{\mbox{\small int}}$ is constant
on $\chi_{0}$-orbit. Also it should not commute with $D_{4}'$,
otherwise it would have the periodic properties, 
which select too narrow class of Hamiltonians. It was shown in paper 
\cite{straight-em1} that not Hamiltonian itself
enters to the physical vertex operators, but its average over a period
of evolution. This averaging actually inserts a projector 
onto the subspace $\chi_{0}=0$. In a similar way we can construct 
an operator $J$, commuting with $\chi_{0}$ or $D_{4}'$. 
Let's consider two cases:

(Ja) $J$ does not contain $X$. In this case $[J,\chi_{0}]=0$
implies $[J,M^{2}]=0$, i.e. $J$ cannot create transitions
between the states with different masses. Also, $[J,D_{4}']=0$
implies $[J,D_{4}]=0$, i.e. $J$ cannot create transitions 
between the states with different $Q$.

(Jb) $J$ contains $X$. In this case the operator 
with $[J,\chi_{0}]=0$ generates simultaneous
changes of $P^{2}$ and $M^{2}$, preserving the mass shell condition. 
In a similar way, $J$ creates transitions between $Q$-sectors,
even if the condition $[J,D_{4}']=0$ is satisfied.

Note that the operator $M^{2}$ itself belongs to
the class (Ja), as a result, it is diagonal
with respect to the quantum number $Q$. 
The vertex operators generally belong to the class (Jb):
they contain $X$-terms, responsible for the change of $P^{2}$. 
E.g. the interaction with a planar wave \cite{straight-em1}
has a factor $e^{ikX}$, changing $P\to P-k$. 
For such interactions $Q$-sectors are mixed up,
therefore no selection rule for $Q$ is necessary.

After factorization of $\chi_{0}$-transformations 
the remaining discrete gauge symmetry is related with the choice 
between different $\chi_{0}$-orbits. For nearly straight string 
they are represented by equivalent solutions in the vicinity 
of the northern pole, the equator, and the southern pole.
In the obtained spin-mass spectrum
the first Regge trajectory corresponds to the 
northern pole solution, further trajectories are created
by expansion (\ref{a2cl}) in the vicinity of the northern pole solution.
Classically there are gauge equivalent solutions near the southern pole
and the equator, which possess the same $(P^{2},S)$. In quantum
mechanics we do not see these additional solutions in the spectrum.
Particularly, the first Regge trajectory is non-degenerate.
We conclude, that the equivalence between these solutions is lost
in quantum mechanics. The spectra for additional solutions
can be shifted to the region of large masses, or the quantum expansions
(\ref{alp2q}) can even diverge on these solutions. Indeed, the classical 
solutions in the vicinity of the northern pole possess
large $n_{1}$ and small $n_{k}$, $k\in\Z/\{0,1,\pm2\}$,
providing the convergence for $(1/n_{1})$-expansion (\ref{a2cl}).
In quantum mechanics the convergence of expansion (\ref{alp2q})
is supported by the normal ordering of operators and
the finite number of occupied modes in $(L_{0}^{(2)}\leq N)$-spaces.
For the solutions near the southern pole $n_{1}\to0$, and the usage
of $(1/n_{1})$-expansions is problematic. One can use $(1/n_{-1})$-series
to construct a definition of mass shell condition, alternative to
(\ref{P2q}), however these definitions will substantially differ 
on the quantum level and in fact will create two distinct theories.
For the solutions on the equator infinite number of oscillator modes
are excited, and for these solutions the convergence of the expansions
is not guaranteed neither on classical nor on quantum level.
This argumentation explains why the usage of $(1/n_{1})$-series
in the vicinity of the northern pole solution preserves only one
Gribov's copy in the quantum theory.

\begin{figure}\label{fcyl}
\begin{center}
~\epsfxsize=6cm\epsfysize=2cm\epsffile{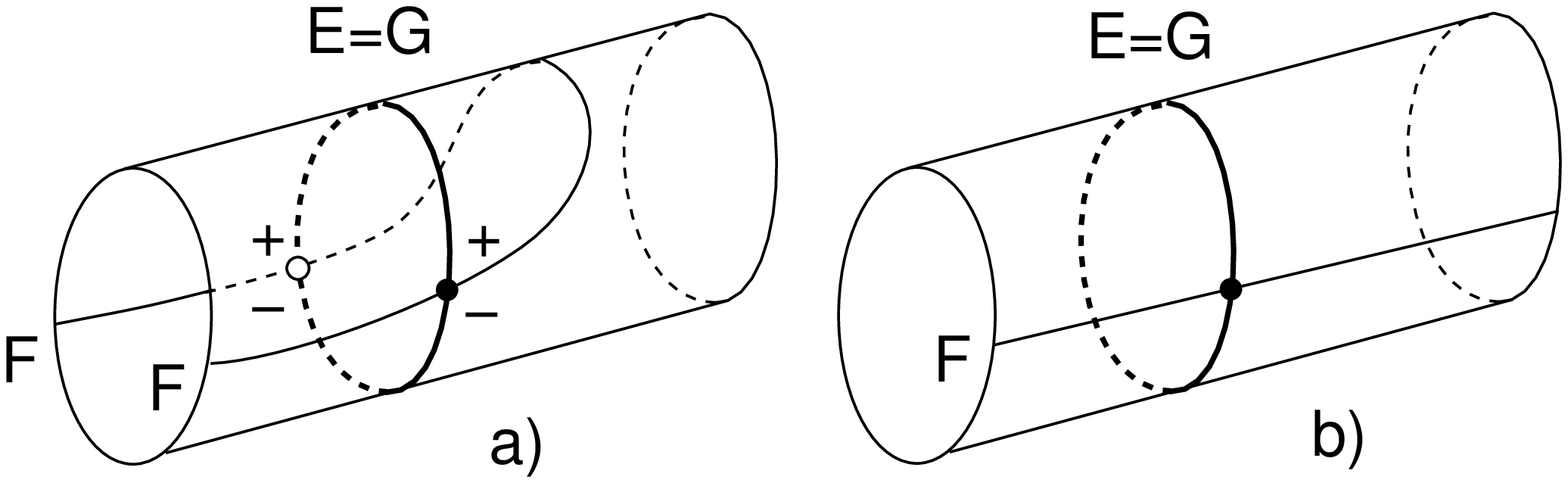}

\fignum Example: Gribov's copies in the phase space\\ 
of planar rotator.

\end{center}
\end{figure}

\noindent{\it Remark.} The problem of Gribov's copies 
in full generality was addressed in the paper \cite{nogauge}.
This paper has considered several systems, possessing 
the Gribov's copies, and discussed various approaches
to their quantization. In the simplest example, planar rotator,
the phase space $M=S^{1}\times\R$ is formed by a periodical coordinate 
$\varphi \mod2\pi$ and spin variable $S\in\R$. There is a constraint 
$E=S^{1}$ of a form $S=Const$, whose gauge group $G=S^{1}$ 
identifies all points on the circle and therefore eliminates 
from the theory all degrees of freedom. 
According to \cite{nogauge}, such system does not
allow gauge fixing condition $F$ of the form $f=0$,
where $f$ is a smooth real-valued function, {\it globally defined} on $M$,
which has only non-degenerate zeros. Indeed, non-degenerate zero of $f$ 
divides the circle to the parts with $f>0$ and $f<0$, 
and there is at least one other point with $f=0$ on the circle, 
see \fref{fcyl}a. However, for construction of the reduced
phase space it is not necessary for $f$ to be globally defined.
For the computation of Dirac's brackets it's only needed that
$f$ has been defined locally, in the vicinity of the intersection 
$E\cap F$. Generally $F$ can be an arbitrary smooth submanifold of $M$,
not necessary representable as zero level of a globally defined function,
see \fref{fcyl}b. Moreover, for the technique \cite{Arnold-mat-phys,Razumov} 
only the intersection $E\cap F$ should be specified and the symplectic 
form of $M$ can be reduced directly to $E\cap F$. 
In the considered example such class of gauge fixing conditions 
allows unique intersection between $E$ and $F$. 
In this low-dimensional example the  
surface of constraint $E$ coincides with the orbit of gauge group $G$,
generally they differ. In certain cases the topology 
of gauge orbit $G$ and gauge fixing condition can lead to multiple 
intersections. Particularly, if both are closed submanifolds of 
Euclidean space (with $dim(G)=codim(F)$), their intersection index 
equals zero, giving an even number of transversal intersections \cite{DNF}. 
The same property holds if one is a closed submanifold and the other one 
is a linear subspace. The set $E\cap F$ can be connected 
(otherwise the connected component can be taken), 
and the orbits $G$ of gauge group, to which $E$ is foliated, 
can intersect $F$ several times. In this case the mechanics possesses 
Gribov's copies and associated discrete gauge symmetries.

\vsp\paragraph*{Appendix 4:} numerical methods.

\begin{figure}\label{fcondense}
\begin{center}
~\epsfxsize=8cm\epsfysize=5.3cm\epsffile{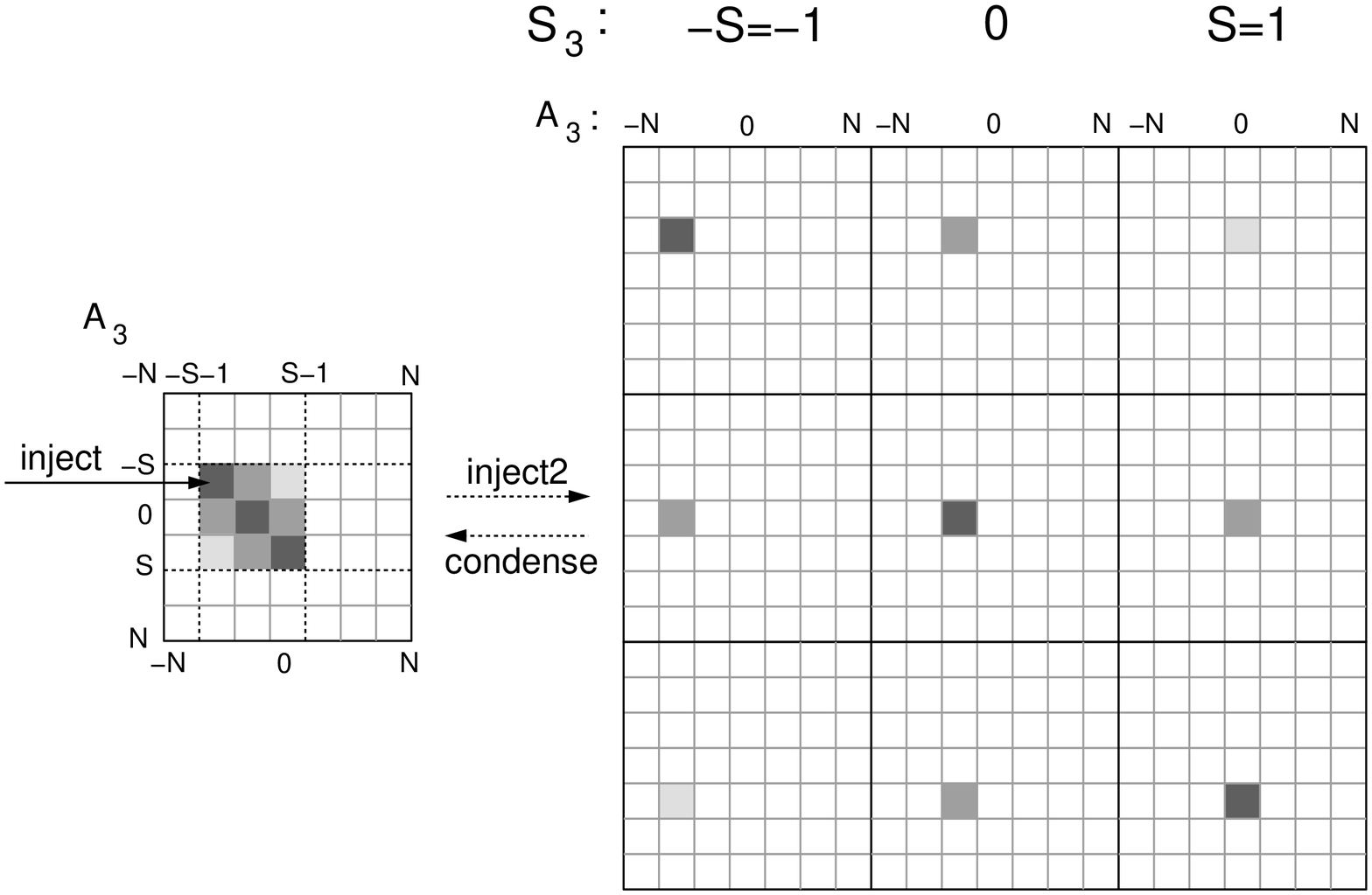}

\fignum $S_{3}$-extension of the algorithm.
\end{center}
\end{figure}

The necessary extensions of the algorithms developed in Part~II
from $S=0$ to $S>0$ consist in the introduction of the additional
block structure, related with the quantum number $S_{3}\in[-S,S]$.
Now we need to compute the matrix elements 
$\bra{\chi_{3}=0,Q}\alpha_{2}\ket{\chi_{3}=1,Q}$,
shown on \fref{fcondense} at the right. Here the large blocks
correspond to $S_{3}$ number, small blocks correspond to $A_{3}$.
The required matrix elements are localized in the intersection 
of rows with $-S\leq S_{3}=A_{3}\leq S$ and columns with 
$-S\leq S_{3}=A_{3}+1\leq S$. Actually, the only extension
is required in the algorithm {\it ord}, procedure {\tt inject},
see Part~II. The results of sparse matrix multiplication
after injection to $\bra{A_{3f},Q}\alpha_{2}\ket{A_{3i},Q}$
blocks in the left matrix on \fref{fcondense} should be further
passed to the right matrix. However, keeping in this matrix 
only necessary elements, we condense it back to the smaller 
left matrix and perform the actual computations in it.
Each monomial has a definite $\Delta A_{3}$-charge and therefore
contributes to one of the diagonals in this matrix, marked on
\fref{fcondense}:
$\bra{S_{3}=A_{3}}mon\cdot S_{+}^{n(S_{+})}S_{-}^{n(S_{-})}
\ket{S_{3}-n(S_{-})+n(S_{+}),A_{3}-n(S_{-})+n(S_{+})-1}
=\bra{A_{3}}mon\ket{A_{3}-n(S_{-})+n(S_{+})-1}\cdot
\bra{S_{3}-n(S_{-})+n(S_{+})}S_{+}^{n(S_{-})}S_{-}^{n(S_{+})}
\ket{S_{3}}.$
The last matrix element gives coefficient
$\prod_{S_{3}-n(S_{-})+n(S_{+})<m\leq S_{3}+n(S_{+})}
\sqrt{S(S+1)-m(m-1)}\cdot\prod_{S_{3}\leq m<S_{3}+n(S_{+})}
\sqrt{S(S+1)-m(m+1)}.$
Here in the case of empty set of indices (e.g. $n(S_{+})=0$)
we define $\prod=1$. If $S_{3}+n(S_{+})>S$ or $S_{3}-n(S_{-})+n(S_{+})<-S$,
we have $\prod=0$. Substituting $S_{3}=A_{3}$ in these formulae,
we have non-zero coefficients only for the values
$A_{3}+n(S_{+})\leq S$, $A_{3}-n(S_{-})+n(S_{+})\geq-S$,
i.e. $A_{3}$ changed between $A_{3min}=\max\{-S,-S+n(S_{-})-n(S_{+})\}$
and $A_{3max}=S-n(S_{+})$. If $A_{3min}>A_{3max}$, the monomial is omitted.
Then we initialize the sparse matrix multiplicator 
as projector to $(A_{3}\in [A_{3min},A_{3max}],L_{0}^{(2)}\mod 4=Q)$ 
subspace and perform the rest computation as described in Part~II.
As an additional optimization, in the products $\prod_{i} op_{i}^{k_{i}}$
we have implemented preliminary computation and storage of the powers 
$op_{i}^{k}$ for $k=1...kmax_{i}$, accelerating this procedure by 
a factor $1.8$.

\begin{figure}\label{ustakan1}
\begin{center}
~\epsfxsize=6cm\epsfysize=4.5cm\epsffile{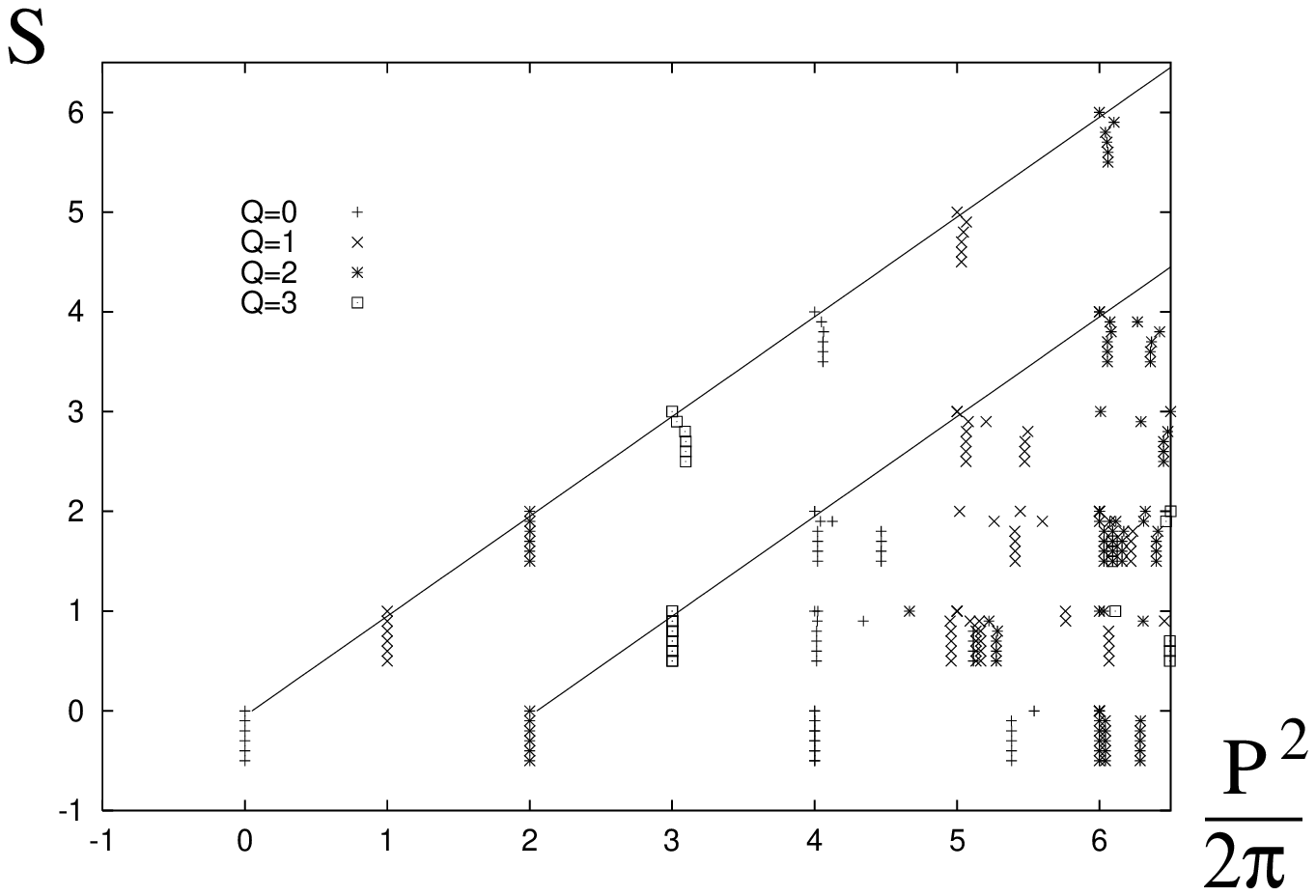}

\fignum Spectrum $(P^{2}/2\pi,S-\epsilon n)$ as a function of $n$.
\end{center}
\end{figure}

\begin{figure}\label{ustakan2}
\begin{center}
~\epsfxsize=6cm\epsfysize=4.5cm\epsffile{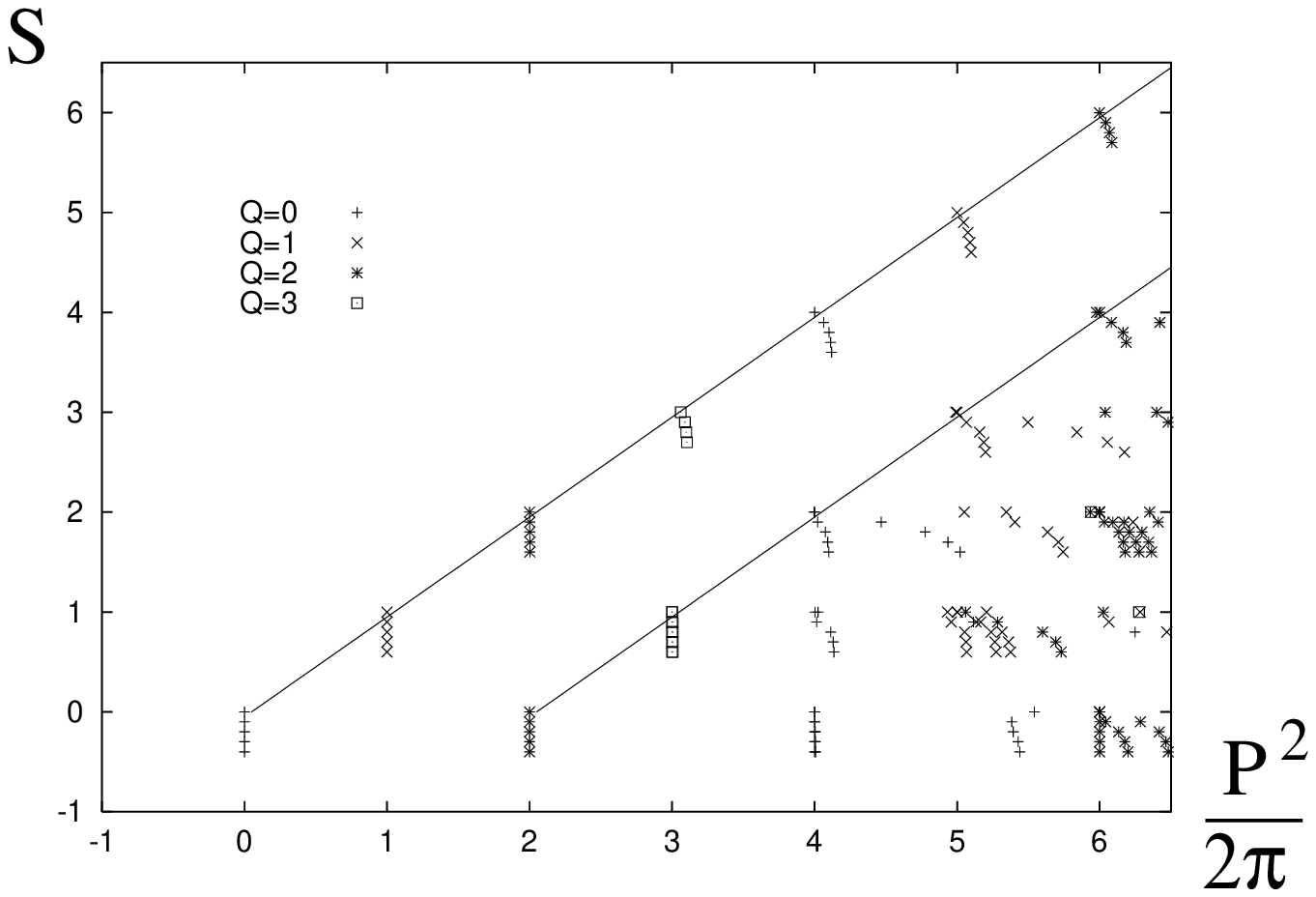}

\fignum Spectrum $(P^{2}/2\pi,S-\epsilon N)$ as a function of $N$.
\end{center}
\end{figure}

The determined spin-mass spectrum
as a function of the correction order $n$ 
and cutoff parameter $N$ is presented on figs 
\ref{ustakan1},\ref{ustakan2}.
The figure \ref{ustakan1}
shows superimposed spectra $(P^{2}/2\pi,S-\epsilon n)$
at fixed $N=11$, where $\epsilon$ is a small constant
and $n$ is changed from $1$ to $6$ in the direction 
from top to bottom. By the reasons explained in the lemmas above, 
some of the states remain fixed (vertical) with the change of $n$,
while the others are rapidly stabilized with $n$. 
On the figure \ref{ustakan2} the spectra $(P^{2}/2\pi,S-\epsilon N)$
are shown for $n=3$, $N=7...20$, where in each $Q$-sector values
$N\mod4=Q$ were selected. When $N$ increases, the most of the states 
also show tendency to stabilization. The spectrum, presented 
on \fref{spec}, is computed for $n=3$ and $N=17,18,19,20$
for 4 values of $Q$ superimposed in this spectrum.

The other statistics of the algorithm is presented
in the following tables:

\footnotesize

\begin{center}
Number of secondary monomials in each order $n=1..6$

\begin{tabular}{|c|cccccc|}\hline
$n$&1  &2 &3 &4 &5 &6\\\hline
num. of monomials&2&8&48&248&1066&3844\\\hline
\end{tabular}
\end{center}

\normalsize

\footnotesize 

\begin{center}
Time and memory requirements of the algorithm $(n=3,S=6,Q=0)$:

\begin{tabular}{|c|cccc|}\hline
$N$ &8  &12 &16 &20 \\\hline
$subdim$ &85$\times$83 &447$\times$438 &1932$\times$1900 &7288$\times$7183 
\\\hline
req.memory &2.1Mb &5.6Mb &83Mb &1.1Gb\\\hline
comp.time &0.08~sec &1.4~sec &14~min &15~hours\\\hline
\end{tabular}
\end{center}

\normalsize

\vspace{3mm} 
In the last table $subdim$ is the dimension of $S=6,Q=0$ block
in dense matrices, which at large $N$ mainly defines the amount 
of required memory. The computation is performed on 
HP 2GHz Linux PC.

\end{document}